\documentclass[review,number,sort&compress]{elsarticle}

\DeclareMathAlphabet{\mathpzc}{OT1}{pzc}{m}{it}
\usepackage{setspace}
\usepackage{epstopdf}
\usepackage{graphicx}
\usepackage{amssymb}
\usepackage{amsmath}
\usepackage{amsthm}
\usepackage[colorlinks=true]{hyperref}
\usepackage{breakurl}
\usepackage[all]{hypcap}
\usepackage{mathrsfs}

\usepackage[percent]{overpic}
\usepackage{fixltx2e}
\usepackage{epsfig}
\usepackage{verbatim}
\usepackage{multirow}
\usepackage{float}
\usepackage{array}

\journal{Nuclear Instruments and Methods A}
\begin{document}
\begin{frontmatter}
\title{Studies of the delayed discharge propagation in the Gas Electron Multiplier (GEM)}

\author[a]{A.~Utrobicic \corref{cor1}}
\ead{antonija@phy.hr}
\author[b]{M.~Kovacic}
\author[a]{F.~Erhardt}
\author[a]{M.~Jercic}
\author[a]{N.~Poljak}
\author[a]{M.~Planinic}

\address[a]{Faculty of Science, University of Zagreb, Croatia}
\address[b]{Faculty of Electrical Engineering and Computing, University of Zagreb, Croatia}

\cortext[cor1]{Corresponding author. Tel.: +385 1 4605553, fax: +385 1 4680336}

\begin{abstract}

This paper presents an investigation of the discharge propagation (DP) to the readout electrode that occurs with a microsecond time delay after a primary discharge that develops inside a GEM foil hole. A single hole THGEM (THick GEM) foil that enables a controlled discharge position and the induction of primary discharge with an over-voltage in the THGEM foil has been used in the initial DP measurements. In order to justify the use of a custom-made THGEM foil, additional measurements were made with a standard GEM foil. Correlated optical (with an ordinary SLR and a high-speed camera) and electrical measurements of the delayed DP were made for Ne-CO$_2$-N$_2$ (90-10-5) mixture and with different powering configurations. Measurements show that the delayed DP happens without a drift field, with an inverted induction field, inverted THGEM voltages or an inverted drift field. After the primary discharge, there is a charge transfer in the induction region at an induction field value below that of the onset field for DP. In the time between the primary discharge and the delayed DP, three different current regimes are observed, which suggests multiple charge transfer mechanisms in the induction region. High-speed camera recordings provide valuable insight into the time evolution of the primary and the delayed DP, especially when correlated with electrical measurements. 

\end{abstract}

\end{frontmatter}

\newpage

\section{Introduction}
\label{sec:chap1}

Modern accelerator facilities such as the LHC and RHIC have set high requirements for gaseous detectors in terms of their radiation hardness, higher rate capability, and time and position resolutions, which resulted in the development of Micro-Pattern Gaseous Detectors (MPGDs). The GEM detector is a type of an MPGD that uses one or more GEM foils in a cascade for charge amplification of electrons created in gas \cite{sauli1997gem, bachmann2002discharge}. Because of their excellent properties, GEM detectors are often used in modern physics experiments and are currently considered as the best candidate for an upgrade of existing gaseous detectors \cite{ketzer2004performance, bagliesi2010totem, appelshaeuser2013technical,abbaneo2013gem}. They have a broad field of usage other than in physics, such as medicine or homeland security.\cite{tsyganov2008gas, gutierrez2012mpgd, aza2017preliminary, gnanvo2011imaging}.

A standard GEM foil consists of a dielectric material (polyimide foil 50 $\mu$m thick) that has a copper layer on both sides (each 5 $\mu$m thick) and has a large number of double conical holes (inner diameter 50 $\mu$m and outer diameter 70 $\mu$m) extending through the foil. Applying a voltage between the upper and the lower copper layers of the foil, high electric fields ($\approx\,$40 kV/cm) are formed in the holes and act as
electron multiplication channels for the charge released by ionization in the gas \cite{bouclier1997new}. The operation of GEM foils at high gains, high rates, and highly ionizing particles can lead to the formation of electrical discharges inside the GEM foil holes. These primary GEM discharges can result in propagating discharges to the readout board, which can be destructive and appear with a microsecond scale time delay after primary discharge  \cite{wallmark2001operating, iacobaeus2001novel, zeuner2000msgc}. 
The probability of the appearance of a delayed DP to the readout board strongly depends on the energy of the primary discharge and on the induction field strength \cite{bachmann2002discharge}. Recent measurements show that this probability has a steep onset from zero to one in dependence on the induction field strength. The time delay between the primary discharge in the GEM foil and the delayed DP to the readout board is reduced with increasing induction field strength. When adding a decoupling resistor between the HV supply and the GEM bottom electrode, the delayed DP probability onset happens at a higher induction field strength \cite{deisting2017discharge}.
An explanation of the time delay in the DP to the readout board is not straightforward since there is no amplification in the GEM holes after the initial discharge. Multiple physical mechanisms have been suggested by several research groups, but none of them can completely explain the nature of the delay \citep{deisting2017discharge,iacobaeus2001novel, wallmark2001operating}.

The main idea of this work was to investigate the delayed DP in order to gain a better understanding of the physical mechanism of its occurrence.  This can help to design additional modifications of GEM detector systems that can reduce the probability of these events and thus expand the detectors' operational range and their robustness. 

\section{Experimental setup}
\label{sec:chap2}

\begin{figure*}[!htb]
\begin{center}
\includegraphics[width=0.99
\columnwidth]{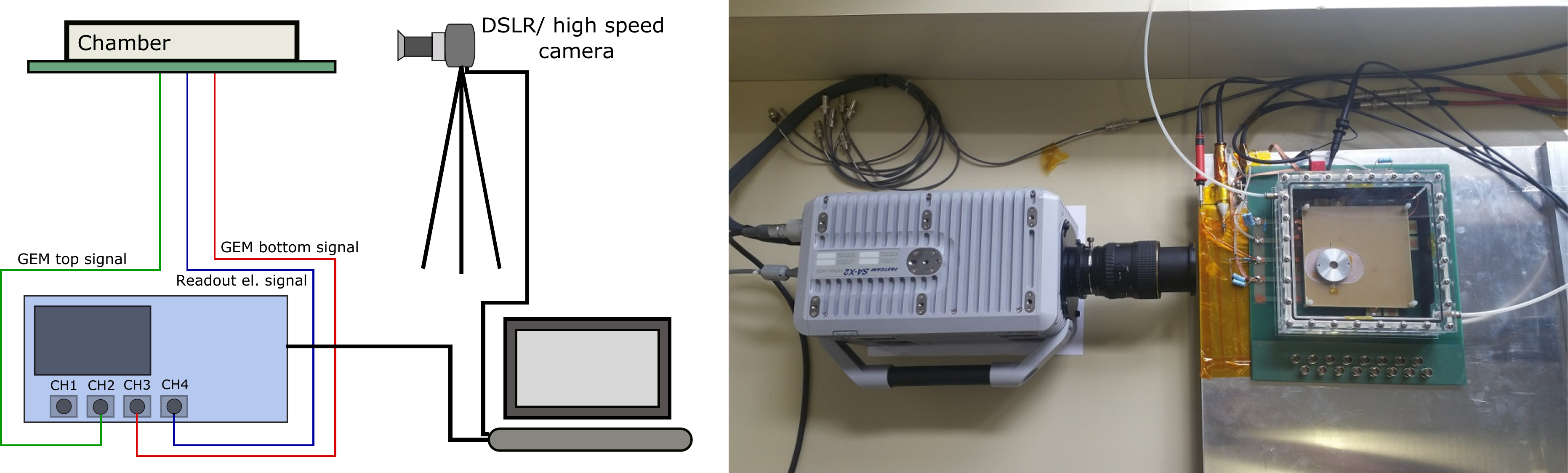}\hspace{0.1 cm}

\caption{Experimental set-up schematics (left panel) and photo (right panel).}
\label{fig_experimental}
\end{center}
\end{figure*}

The measurements were made with a custom-made transparent acrylic glass chamber where drift and a copper plate readout electrode were placed together with either a single hole THGEM or a standard GEM foil. To be able to control the position of the primary discharge,  a  drift electrode with a small drilled hole was used. When performing measurements with a THGEM foil and a normal induction field orientation, a drift electrode was not used. Primary discharges were induced by applying an over-voltage to the THGEM electrodes. Either an SLR or a high-speed camera was used to record primaries and the delayed discharges. The camera was triggered by an oscilloscope which was set to trigger on the primary discharges. The oscilloscope signals from the (TH)GEM top electrode, (TH)GEM bottom electrode or the readout electrode were recorded simultaneously with optical measurements by a computer.  The schematics and photo of the experimental set-up are shown in figure \ref{fig_experimental}.

Three different oscilloscope probes were used to record the signals from the electrodes. The relatively low input resistance and a high input capacitance of commercial HV probes makes them inappropriate for measurements on the top GEM electrode supplied with an M$\Omega$ resistance. Due to the high resistance of the burden resistor required to power (TH)GEM top electrode, a special probe had to be constructed. This probe was constructed using a combination of a capacitive divider and a commercial LeCroy 10x probe. 
The capacitive divider was constructed from a 2~pF coaxial Teflon (PTFE) capacitor in a custom probe. A surface mount 100~pF capacitor was placed in parallel to the probe output to make a capacitive divider. By adding the 10x oscilloscope probe, a probe of approximately 500x was constructed.

To measure the (TH)GEM bottom electrode voltage, a commercial LeCroy HV 4~kV (100:1, 400~MHz, 50~M$\Omega$) probe was suitable due to the relatively low resistance of the decoupling resistors ($<$100~k$\Omega$). 

The signal from the readout electrode was connected to GND over a parallel connection of a 100~nF capacitor and a 100~k$\Omega$ resistor and was measured with a standard LeCroy 600~V (10:1, 500~MHz, 10~M$\Omega$) oscilloscope probe. The purpose of this measurement was to determine the current in the induction region between the primary and the delayed DP. The current was obtained from the time derivative of the voltage on the capacitor. A shunt current measurement was not used in order to avoid the introduction of an additional decoupling resistance from the readout side that will reduce the induction field during the delayed DP.

\subsection{Experimental setup with the custom-made THGEM foil}

\begin{figure*}[h t b]
\begin{center}
\includegraphics[width=0.75
\columnwidth]{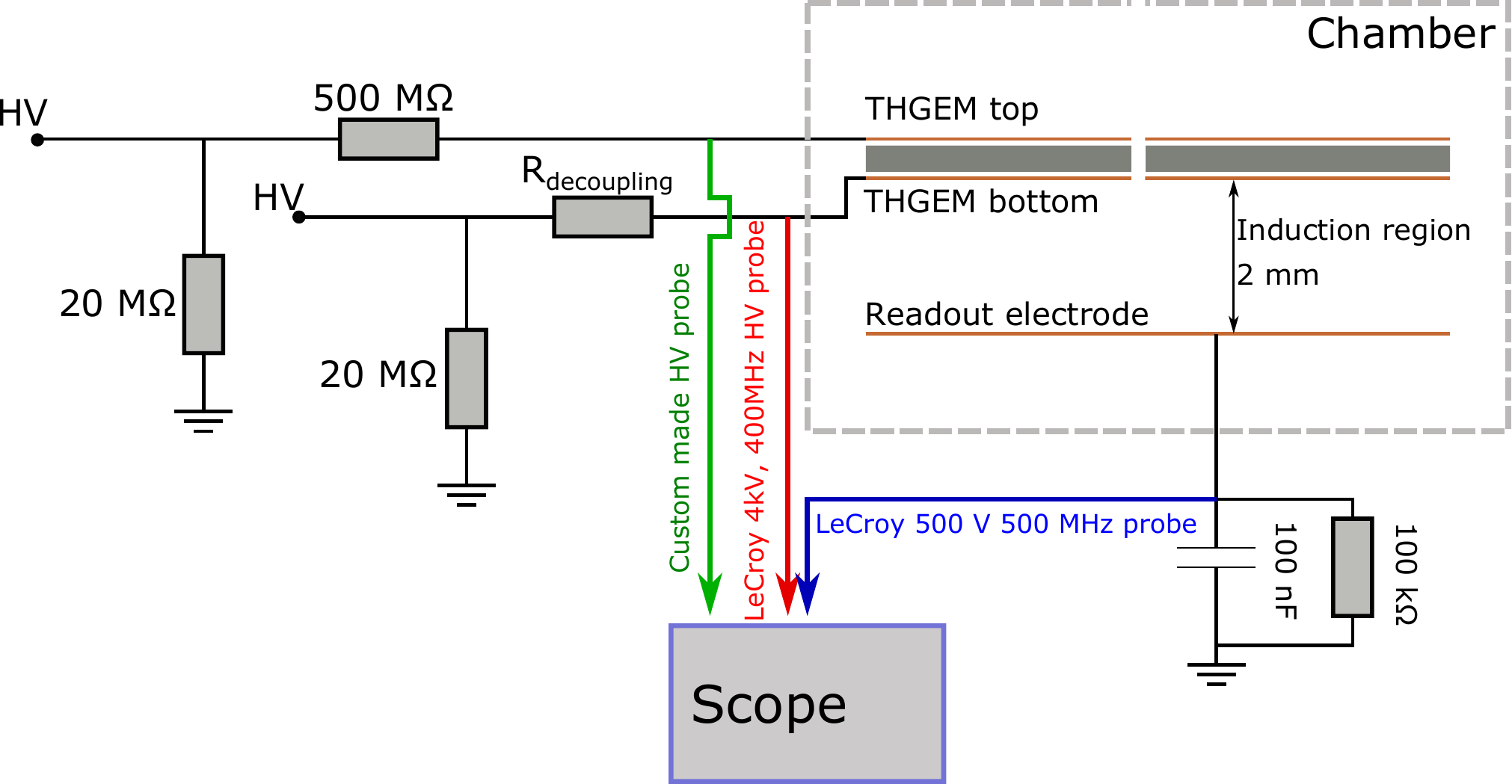}\hspace{0.1 cm}
\caption{Schematics of the THGEM chamber and the powering circuit.}
\label{fig_power_schame_THGEM}
\end{center}
\end{figure*}

The schematics of the THGEM chamber and a powering circuit are shown in figure \ref{fig_power_schame_THGEM}. In order to make a large number of measurements with a known discharge position, a rugged THGEM foil with a single hole was designed and manufactured. The foil was made from a 0.2~mm thick FR4 dielectric material that has a 17.5~$\mu$m thick copper layer with an area of 10 x 10 cm$^2$ on both sides. A single hole, 0.3~mm in diameter, was drilled through the foil in order to control the primary discharge position.

The primary discharge inside the THGEM single hole was induced by applying an over-voltage to the THGEM electrodes. Afterward, the value of the induction field was increased until the DP probability hit 100\%. This method enabled measurements without a drift field and a drift cathode above the THGEM top electrode and without the use of a radioactive source to induce the primary discharges. The measurements were made in Ar-CO$_2$ (70-30), Ne-CO$_2$-N$_2$ (90-10-5) gas mixtures and in the air. Within this paper results obtained with Ne-CO$_2$-N$_2$ mixture are presented.

Both THGEM electrodes are powered through an independent HV channel from an \textit{Iseg EHS8080n} power supply. A loading resistor of 500~M$\Omega$ was used on the THGEM top electrode to limit the number of sparks recorded by the camera to a single spark. A decoupling resistor was used in series with the THGEM bottom electrode. Values of 0~k$\Omega$, 50~k$\Omega$ or 100~k$\Omega$ were used for the decoupling resistors. Since the power supply cannot sink the current, a 20~M$\Omega$ current sinking resistor was connected between the HV outputs and GND.


\subsection{Experimental setup with a standard GEM foil}
\begin{figure*}[h t b]
\begin{center}
\includegraphics[width=0.75
\columnwidth]{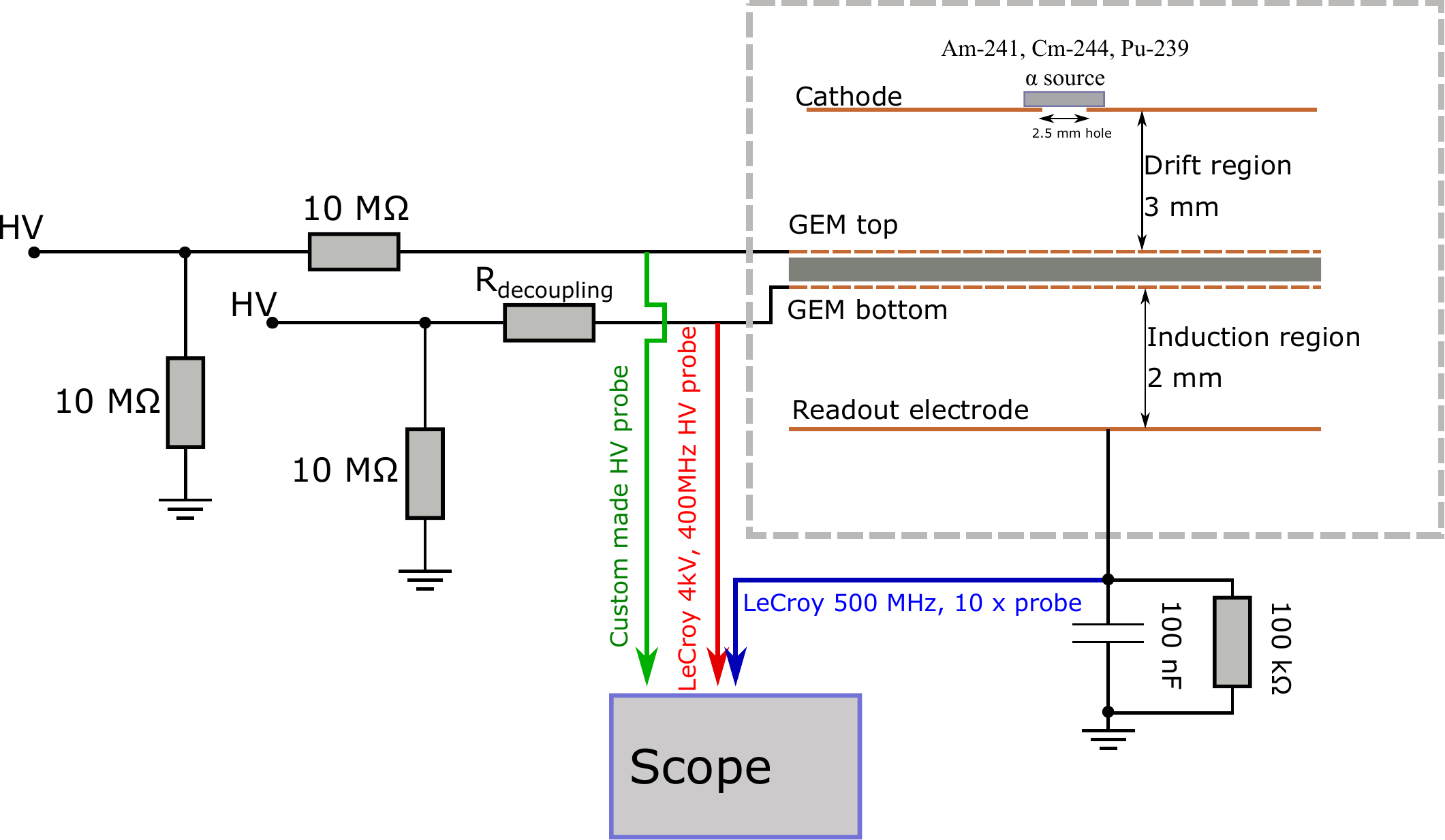}\hspace{0.1 cm}
\caption{Schematics of the GEM chamber and the powering circuit.}
\label{fig_power_schame_GEM}
\end{center}
\end{figure*}

Figure \ref{fig_power_schame_GEM} shows the chamber and powering schematics used for the single stage GEM detector measurements. A large pitch 10 x 10 cm$^2$  GEM foil manufactured at CERN with drift and readout electrode was mounted inside the chamber. An ordinary single sided PCB was used as a drift cathode. A mixed nuclide ($^{241}$Am, $^{244}$Cm and $^{239}$Pu) alpha emitting source was mounted inside the chamber directly on the drift cathode in order to induce a primary discharge in the GEM foil. A hole with a diameter of 2.5~mm  was drilled into the cathode to allow the radiation to enter the drift volume. This enabled rough control of the primary discharge position on the GEM foil within a small area below the drift cathode hole, which was crucial for precise focusing of the camera. All measurements were made in a Ne-CO$_2$-N$_2$ (90-10-5) gas mixture.

Voltages to the drift electrode and GEM electrodes were applied from three independent HV channels. The GEM top electrode was connected over a 10~M$\Omega$ resistor and the GEM bottom electrode over a 0~k$\Omega$, 50~k$\Omega$ or a 100~k$\Omega$ decoupling resistor. A 10~M$\Omega$ current sinking resistor was used on each of the HV channels. The drift field was set to 800~V/cm.

\section{Measurements}
\subsection{Delayed DP measurements with a custom made THGEM foil}

A delayed DP to the readout electrode was observed in Ar-CO$_2$ (70-30), Ne-CO$_2$-N$_2$ (90-10-5) and even in the air, with a custom-made single hole THGEM foil. Delayed DP measurements were made in a way that the induction field value was increased step by step while the voltage difference on the THGEM was kept constant.

\begin{figure*}[h t b]
\begin{center}
\includegraphics[width=0.99
\columnwidth]{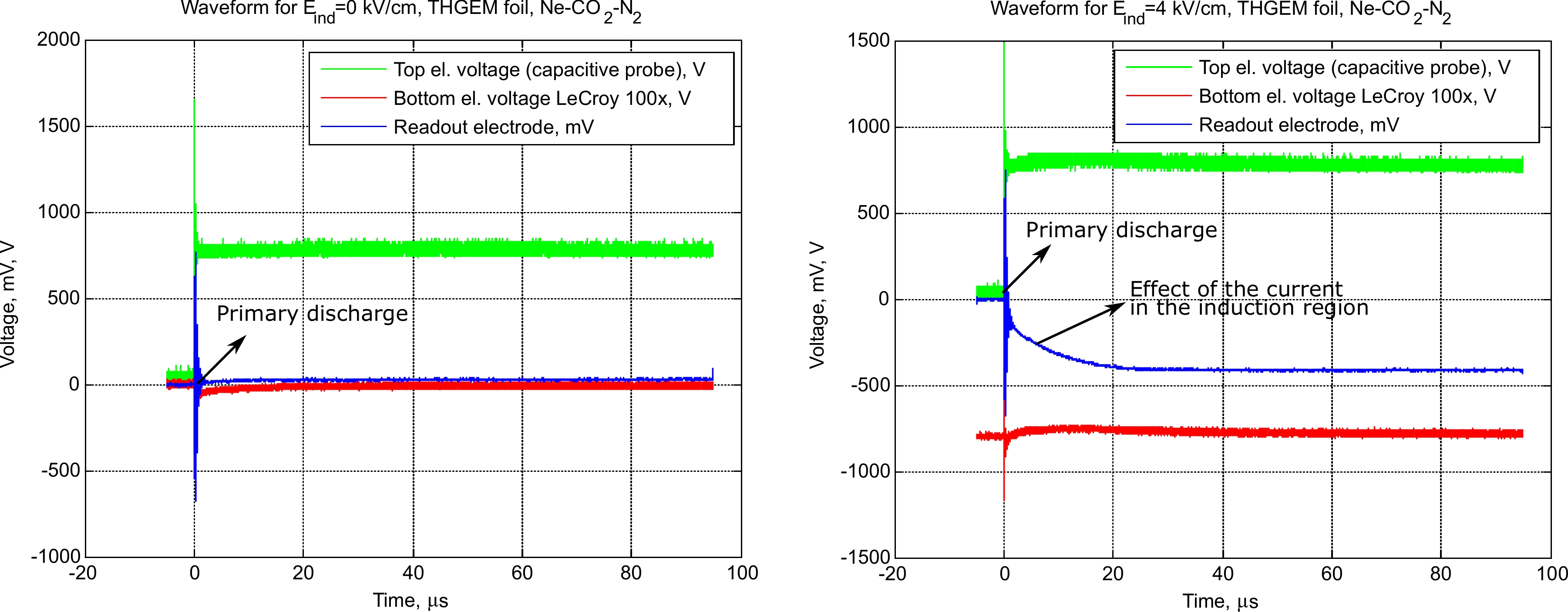}
\caption{Recorded oscilloscope waveforms for measurements with a THGEM foil for induction fields 0~kV/cm and 4~kV/cm in a Ne-CO$_2$-N$_2$ (90-10-5) mixture with $\Delta V_{THGEM}$=900~V.}
\label{fig_4}
\end{center}
\end{figure*}  

Figure \ref{fig_4} shows the time evolution of the recorded voltage signals from the THGEM top and bottom electrodes, and the readout electrode. The green signal represents the voltage on the THGEM top electrode without the DC component. This signal is AC due to the design of the custom-made capacitive divider probe that was used.  The red signal shows the voltage on THGEM bottom electrode that is DC. The blue signal shows the voltage on the 100~nF capacitor connected to the readout electrode. The voltage (y-axis) is shown in two scales. The voltage from the 100~nF capacitor is given in mV and voltages from the THGEM electrodes are given in V.

The left plot on the figure \ref{fig_4} shows the recorded waveforms for measurements at an induction field value of 0~kV/cm and a 900~V potential difference across the THGEM electrodes. On the recorded waveforms the primary discharge can be observed at t= 0~s. To aid visual comparison all the recorded waveforms are plotted in a way that the primary discharge occurs at t=0~s. After the primary discharge, the potential of the top electrode changes from the negative value to the potential of the bottom electrode. The voltage on the capacitor is unchanged after the primary discharge which indicates that at this time  no current is flowing to the readout electrode. 

The right plot on the figure \ref{fig_4} shows the primary discharge for an induction field value of 4 kV/cm. 
A slope can be observed in the recorded voltage on the 100~nF capacitor after the primary discharge. This signal is the integral of the current flowing to the readout board and indicates that even at moderately low values of the induction field there is a mA current in the induction region. 

\begin{figure*}[h t b!]
\begin{center}
\includegraphics[width=0.9
\columnwidth]{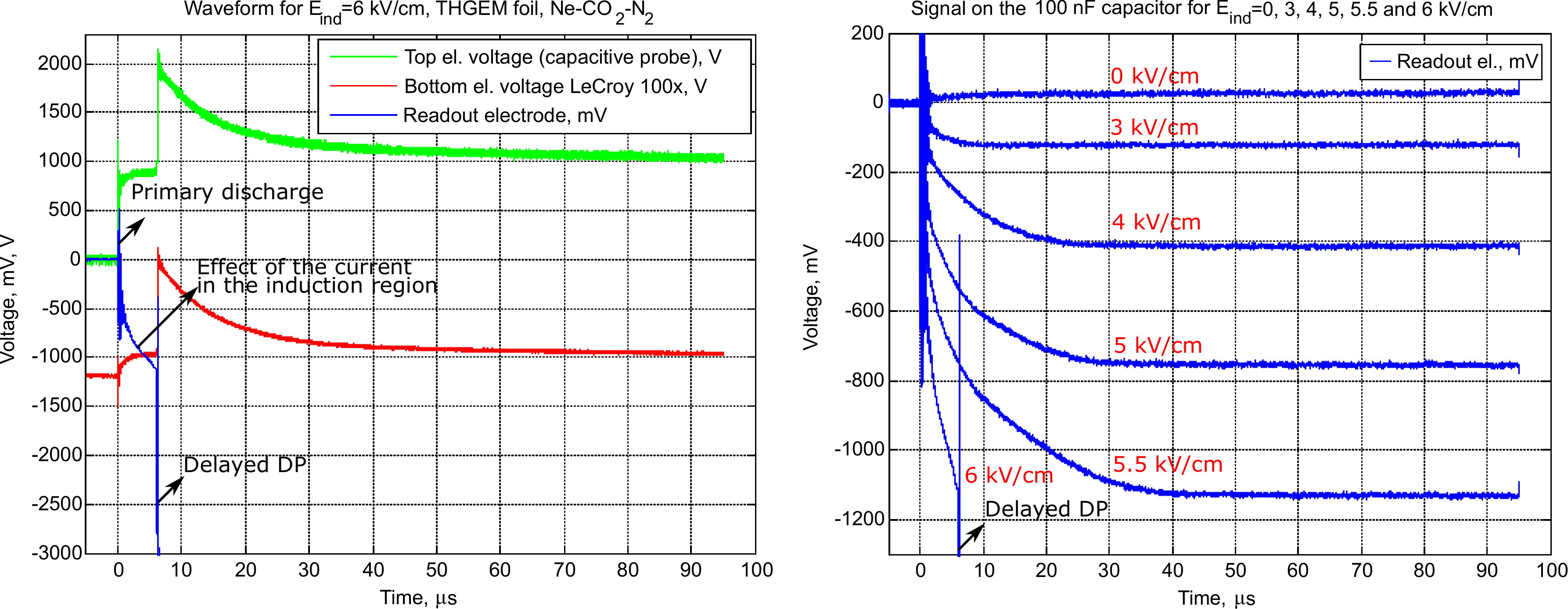}
\caption{Recorded oscilloscope waveforms for measurements with a THGEM foil at an induction field of 6~kV/cm (left) and the recorded voltage on the 100~nF capacitor connected to the readout board (right) at different induction field values in a Ne-CO$_2$-N$_2$ (90-10-5) mixture.}
\label{THGEM_Neon_wf_0k_6kV_cap}
\end{center}
\end{figure*}   

The left plot of figure \ref{THGEM_Neon_wf_0k_6kV_cap} shows measurements at higher induction fields value, where a delayed DP occurs. It can be observed that the delayed DP occurs 6~$\mu s$ after the primary discharge. The recorded voltage from the 100~nF capacitor indicates that in the time period between the primary and the delayed DP there is a charge transfer in the induction region. 

The voltage measurements on the readout capacitor (100~nF) for various induction field values are shown on the right panel of the figure \ref{THGEM_Neon_wf_0k_6kV_cap}. The current in the induction region is increased when increasing the induction field value, which indicates that the charge transfer in the induction region increases significantly with increasing field strength.




Recent research indicates that using a decoupling resistor between the HV supply and the GEM bottom electrode shifts the occurrence of the delayed DP to higher induction field values \citep{deisting2017discharge}. In order to gain more insight into the current in the induction region after primary discharge, simultaneous optical and electrical measurements were made with a 50 k$\Omega$ and without a decoupling resistor. 

\begin{figure*}[h t b!]
\begin{center}
\includegraphics[width=0.9
\columnwidth]{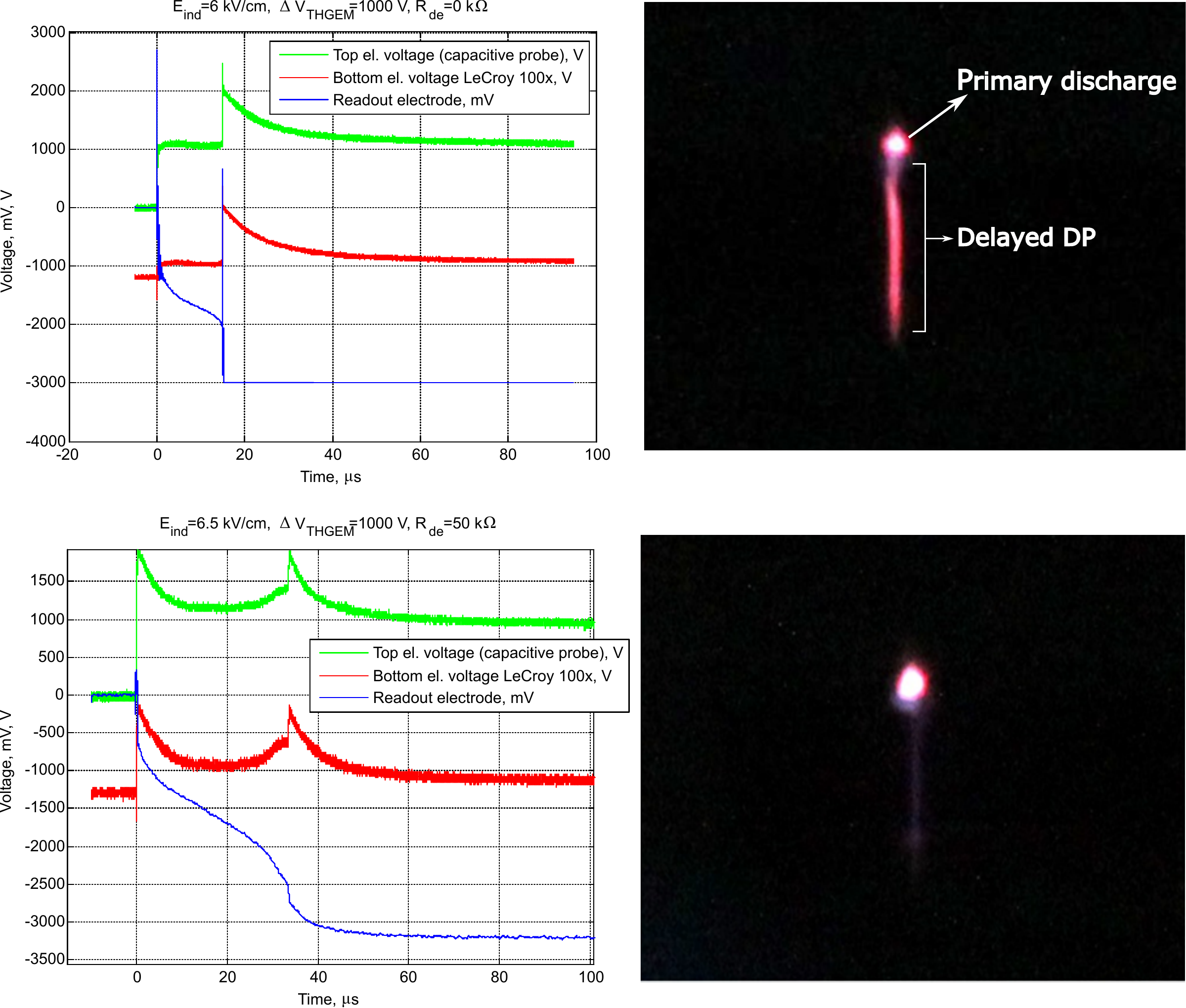}
\caption{Simultaneously recorded oscilloscope waveforms with photographs of primary and the delayed DP without (upper panel) and with 50 k$\Omega$ (bottom panel) decoupling resistor in a Ne-CO$_2$-N$_2$ (90:10:5) mixture.}
\label{fig_6}
\end{center}
\end{figure*}


Figure \ref{fig_6} shows the simultaneously recorded electrical (left panels) and optical (right panels) measurements without (upper panels) and with a 50 k$\Omega$ decoupling resistor (bottom panels). The recorded waveforms suggest that in both cases there is a current in the induction region after the primary discharge that can be observed as a build-up of negative charge on the readout electrode capacitor.  Optical recordings show that the light emitted by the delayed DP is of lower intensity when using a 50 k$\Omega$ decoupling resistor since the decoupling resistor limits the current and quenches the DP. This is also observed on the recorded readout electrode waveforms, showing that the delayed DP slope is smaller in amplitude for measurements with a 50 k$\Omega$ resistor. 

\subsection*{Measurements of the delayed DP with inverted induction field, GEM voltages or drift field}

Various HV powering configurations were tested in order to obtain information on how the inversion of the fields in the THGEM detector affects the delayed DP. Measurements of delayed DP with an inverted induction field, GEM voltages or drift field were made with a custom-made THGEM in a Ne-CO$_2$-N$_2$ (90-10-5) gas mixture. For these measurements, the drift cathode with a hole was added above the THGEM. A primary discharge was induced with a combination of a slight over-voltage on the THGEM and a radiation source.  

\begin{figure*}[!htb!]
\begin{center}
\includegraphics[width=0.9
\columnwidth]{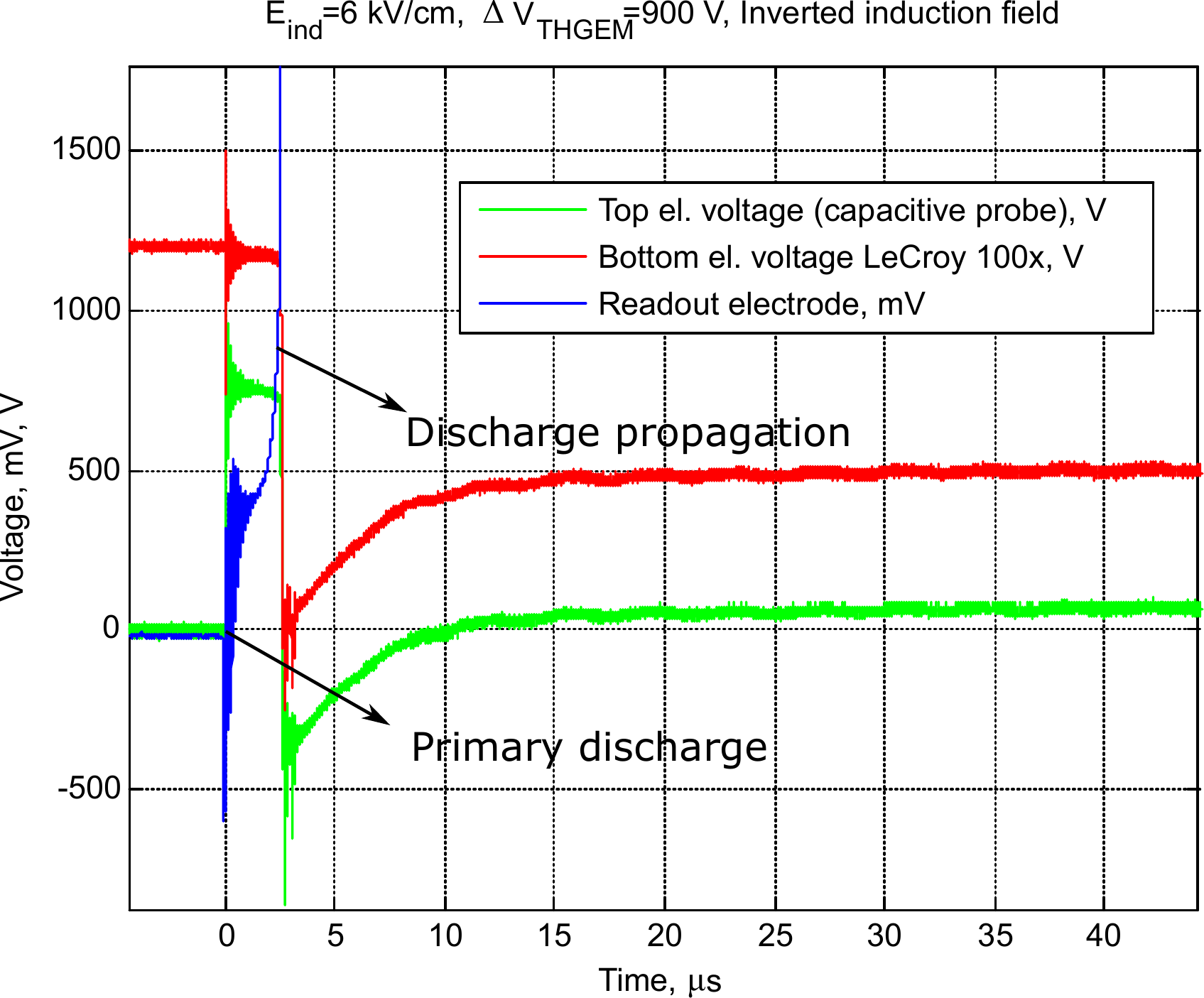}
\caption{Inverted induction field measurements with the custom made THGEM for the induction field value of 6~kV/cm in a Ne-CO$_2$-N$_2$ (90-10-5) gas mixture.}
\label{fig_10}
\end{center}
\end{figure*} 

Firstly, a configuration with an inverted induction field was tested. The readout electrode was connected to GND (through 100 nF capacitor) and a positive voltage was applied on the THGEM bottom electrode, thus inverting the induction field, while the THGEM voltages and the drift field remained in the standard configuration. Figure \ref{fig_10} shows the recorded waveform at an induction field of 6~kV/cm. 2.5~$\mu$s after the primary discharge the delayed DP is observed. If this waveform is compared to the waveform of normal induction field on figure \ref{fig_4}, it can be seen that the inversion of the induction field significantly alters the charge transfer in the induction region right after the primary discharge, but doesn't prevent the delayed DP. 

\begin{figure*}[htb!]
\begin{center}
\includegraphics[width=0.9
\columnwidth]{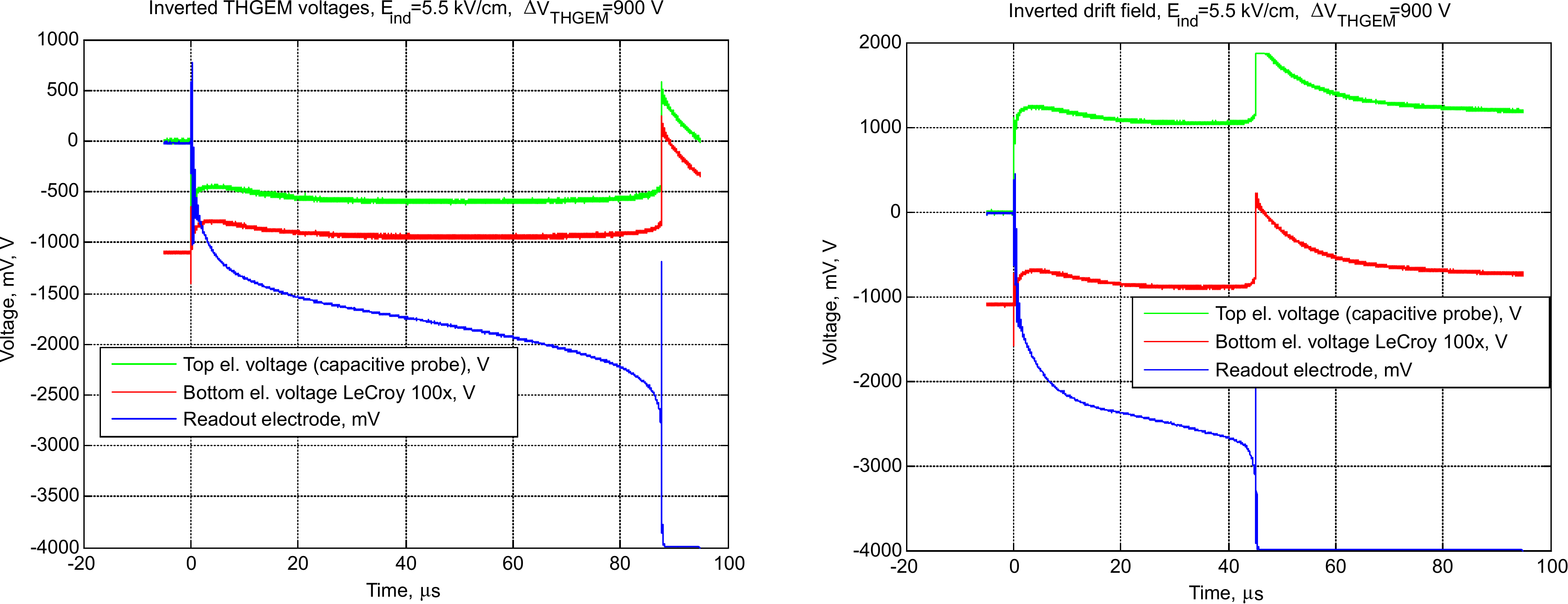}
\caption{The recorded waveforms for measurements with inverted THGEM voltages (left panel) and an inverted drift field (right panel) for an induction field value of a 5.5~kV/cm in a Ne-CO$_2$-N$_2$ (90-10-5) gas mixture.}
\label{fig_11}
\end{center}
\end{figure*} 

Recorded waveforms for measurements with either inverted THGEM voltages or the drift field are shown in figure \ref{fig_11}. A negative voltage slope (on the 100~nF capacitor connected to the readout electrode) was measured in the time between the primary discharge and the delayed DP, similar to the measurements with a normal induction field. This suggests that the charge transfer in the induction region that leads to the delayed DP is not affected by this configuration. 


\subsection{Delayed DP measurements with an LP GEM foil}
The studies performed with a THGEM were repeated with an LP GEM foil in order to see if the same effect of current in the induction region is observed. In this case, the measurements were made only in a Ne-CO$_2$-N$_2$ (90-10-5) gas mixture. 

\begin{figure*}[h t b]
\begin{center}
\includegraphics[width=0.9
\columnwidth]{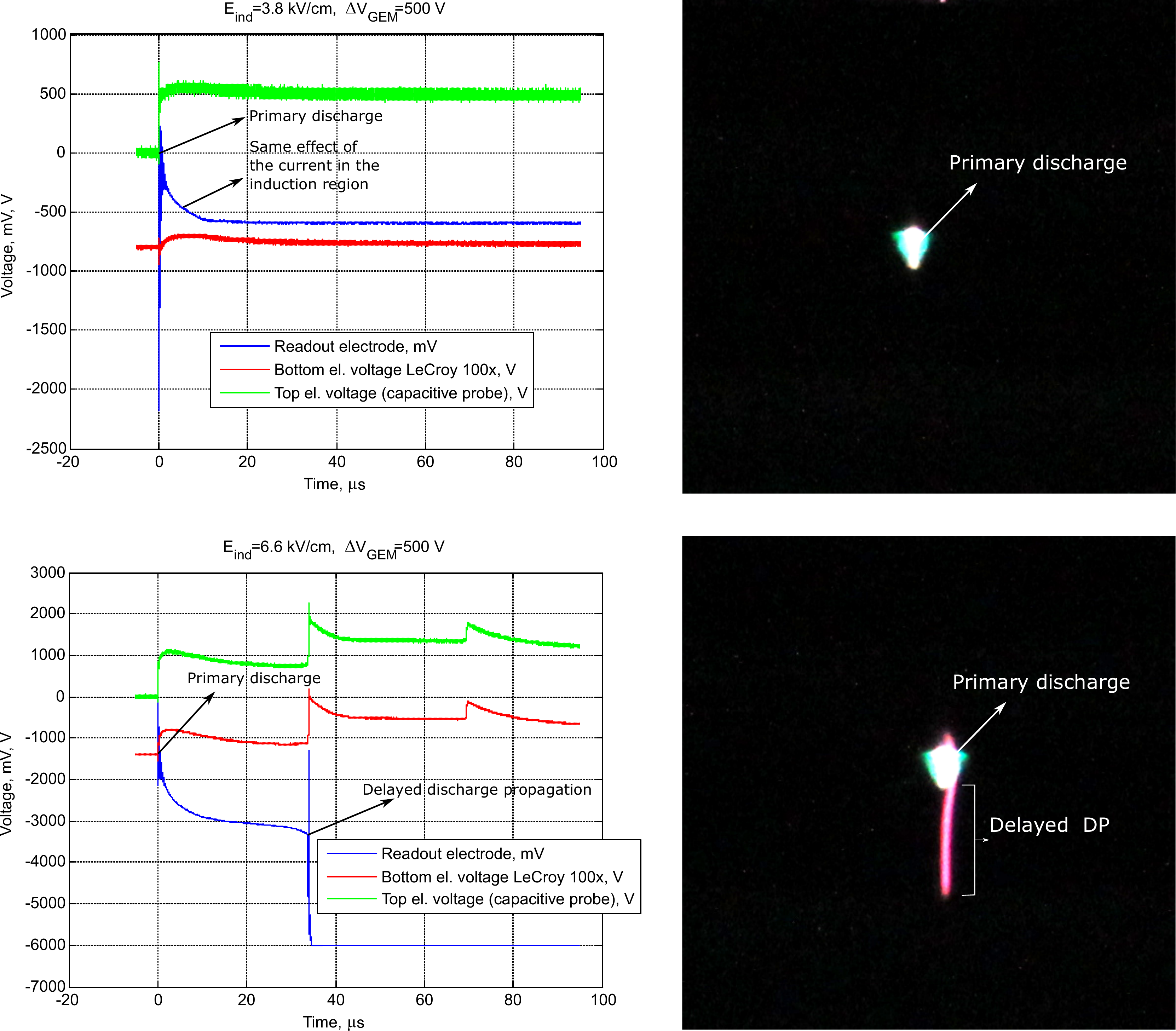}
\caption{Simultaneously recorded waveforms (left) and photograph (right) for induction field value of 3.8~kV/cm (up) and 6.6~kV/cm (down) for measurements with LP GEM foil without decoupling resistor in a Ne-CO$_2$-N$_2$(90-10-5) gas mixture.}
\label{GEM_Ne_0k_2mm_wf_ph}
\end{center}
\end{figure*} 

Figure \ref{GEM_Ne_0k_2mm_wf_ph} shows simultaneously recorded waveforms and photographs for induction field values of 3.8~kV/cm (top panel) and 6.6~kV/cm (bottom panel). The recorded waveform for 3.8~kV/cm shows the same effect of current decay in the induction region after the primary discharge, just as in measurement with a THGEM foil. For a higher induction field value, where the delayed DP occurs, a current in the induction region in the time period between the primary and the propagating discharge can be observed.   

\begin{figure*}[h t b]
\begin{center}
\includegraphics[width=0.9
\columnwidth]{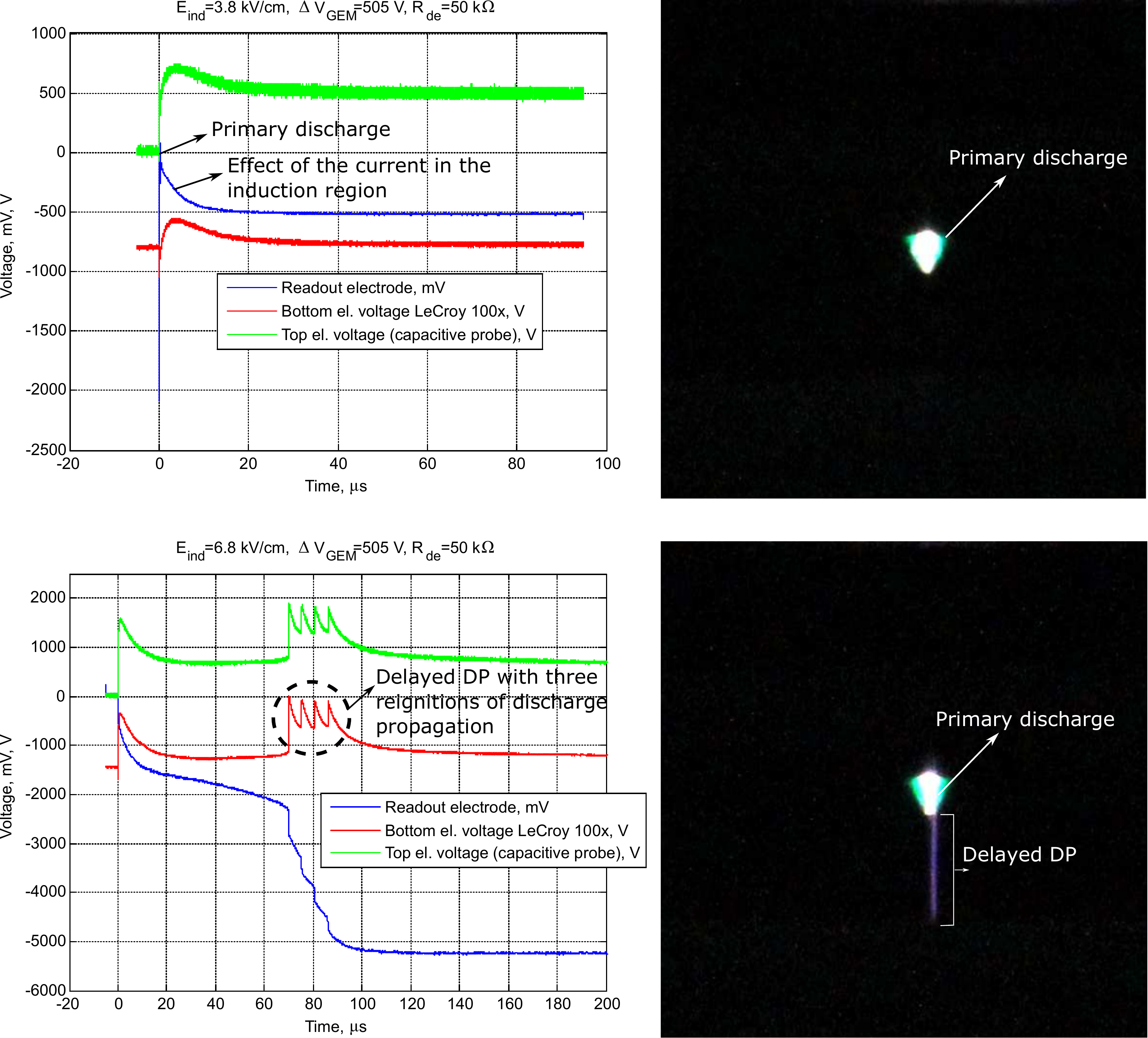}
\caption{The recorded oscilloscope waveforms for measurements with a LP GEM foil for inductions field values of 3.8~kV/cm and 6.8~kV/cm with a 50k~$\Omega$ decoupling resistor in a Ne-CO$_2$-N$_2$ (90-10-5) mixture.}
\label{GEM_Ne_50k_2mm_wf_ph4}
\end{center}
\end{figure*} 

Similarly to the THGEM measurements, a decoupling resistor was added to test its influence on the delayed DP onset field. Electrical (left panels) and optical (right panels) measurements for the induction field values of 3.8~kV/cm (top panel) and 6.8~kV/cm (bottom panel) are shown on figure \ref{GEM_Ne_50k_2mm_wf_ph4} for decoupling resistor value of 50 k$\Omega$. The recorded waveforms suggest that for induction field values lower than that of the DP onset field value, the same effect of current in the induction region is observed. For higher induction fields, when the delayed DP is observed (figure \ref{GEM_Ne_50k_2mm_wf_ph4} bottom panels), a current is observed in the time interval between the primary and the delayed DP. 

From the recorded waveform it can be observed that shortly after the delayed DP, a reignition of the DP can occur. On the bottom left waveform of figure \ref{GEM_Ne_50k_2mm_wf_ph4} three DP reignitions can be observed. The voltage on the GEM bottom electrode starts to rise from zero to approximately  -600~V after the first DP, and then sharply falls back to zero with each reignition. This can not be seen on the photo (bottom right panel) since all of the reignited DP-s are formed along the same path.

\begin{figure*}[h t b]
\begin{center}
\includegraphics[width=0.9
\columnwidth]{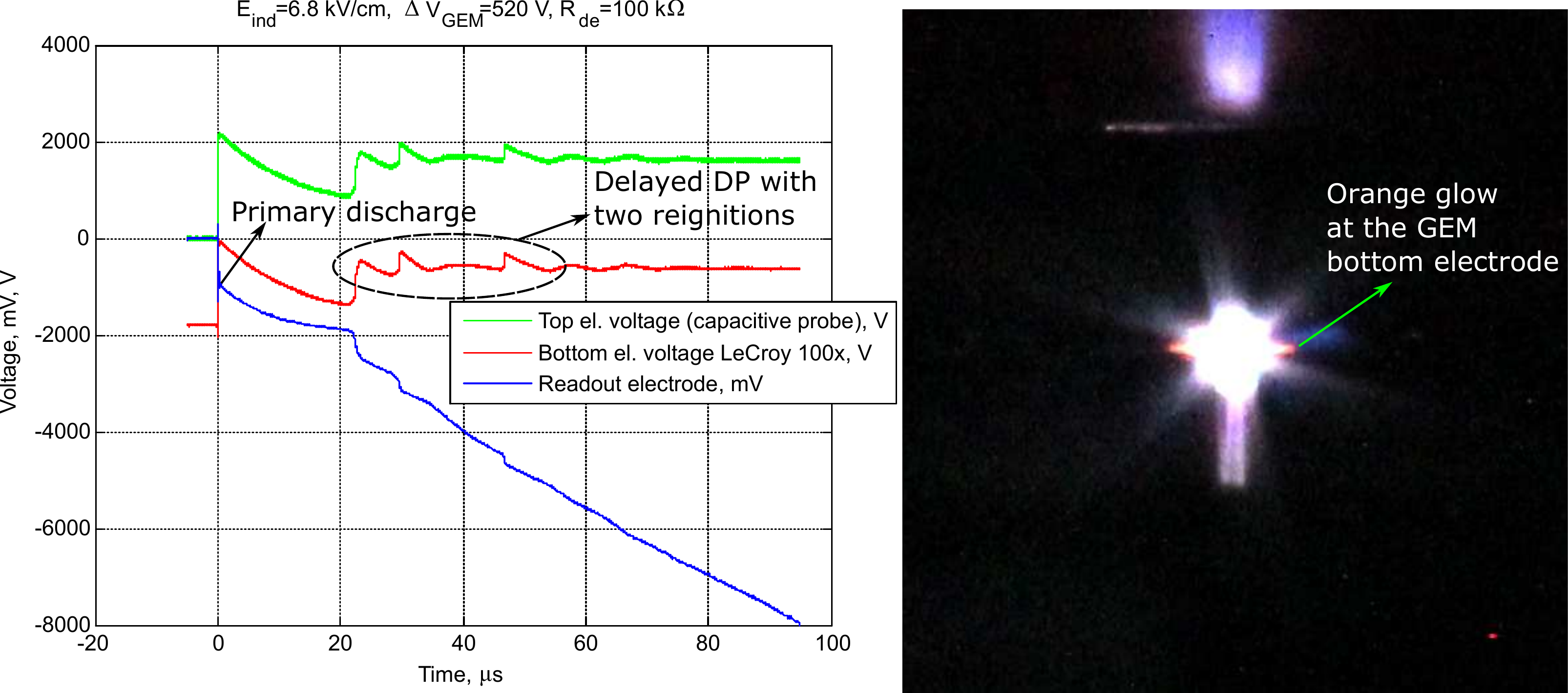}
\caption{The recorded oscilloscope waveforms and photograph for measurements with a LP GEM foil for inductions field values of a 6.8~kV/cm with a 100~k$\Omega$ decoupling resistor in a Ne-CO$_2$-N$_2$ (90-10-5) mixture.}
\label{fig_9}
\end{center}
\end{figure*}

Measurements with a 100 k$\Omega$ decoupling resistor are shown on figure \ref{fig_9}. One can observe that 20~$\mu$s after the primary discharge, there is a propagating discharge followed by two re-ignitions. The voltage on the GEM bottom electrode during the primary discharge drops to zero. This can wrongly suggest that the discharge propagation occurs during primary discharge. The reason why the induction field value drops to nearly 0~kV/cm is due to considerable quenching during the primary discharge induction current peak.

The actual delayed DP occurs  22~$\mu$s after the primary when induction field restores. Afterward, the voltage on the GEM bottom electrode remains at a constant value of -600~V. The voltage on the capacitor connected to the readout electrode suggests that during this time there is a charge transfer in the induction region and a sustained discharge is formed. 

A matching photograph on the right panel on the figure \ref{fig_9} shows two discharge events due to the long time that the camera required to close the shutter after the waveform was recorded by the oscilloscope. The photo shows that the delayed discharge propagation is different in optical appearance when using a 100~k$\Omega$ decoupling resistor than when there is no decoupling resistor. The orange glow visible near the GEM bottom electrode in the vicinity of the primary discharge reveals a possible charge source in the induction region after the primary discharge. The glow could hint at the physical nature of a delayed DP, but it is not always visible on the photos as the position of the glow is very near to the GEM hole and is often hidden by the intense primary discharge. Due to this problem, measurements with an ultra-fast camera were needed to be performed in order to distinguish the primary from the delayed DP and to provide temporal information between those events.

\section{ High-speed camera measurements}

A Photron SA-X2 1080K fast camera was used for optical recordings of the delayed DP. This camera can record at 1080000 fps with a resolution of 8x128 pixels. Based on the geometry of the experimental set-up and the dimensions of the delayed DP, a 100 mm F2.8 macro lens was used for recording. An oscilloscope was used to trigger the camera recording, whose settings were adjusted from a PC. The primary discharge in the THGEM was induced only with an over-voltage while in a standard GEM  a combination of over-voltage and radiation was used.

\subsection{ High-speed camera measurements with a custom-made THGEM foil}
\begin{figure}[!htb]
\begin{center}
\includegraphics[width=1.0
\columnwidth]{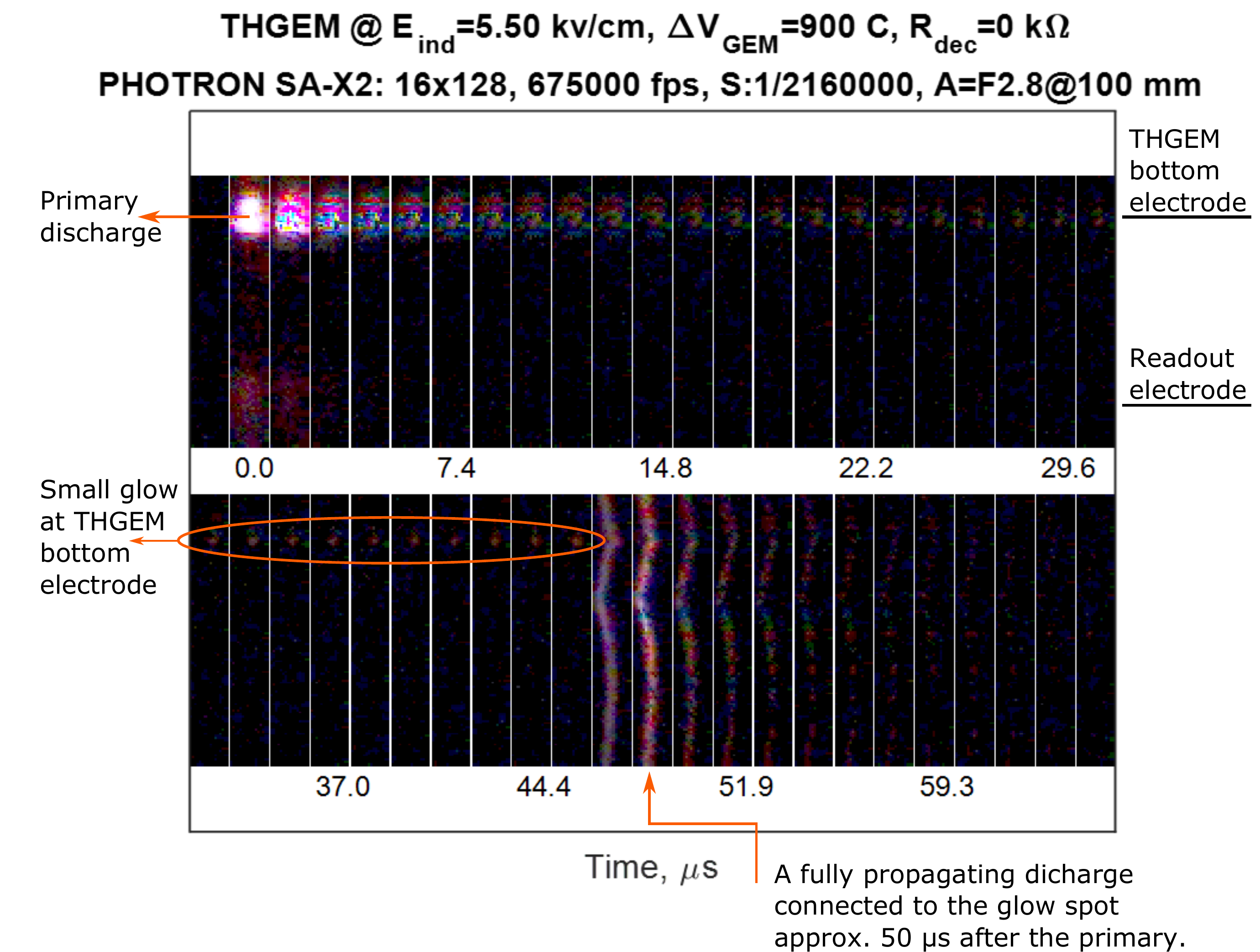}
\caption{A fast camera recording of the delayed DP with the THGEM foil at an induction field value of 5.5 kV/cm without a decoupling resistor in a Ne-CO$_2$-N$_2$ (90-10-5) gas mixture.}
\label{fig_12}
\end{center}
\end{figure}

The first measurements with a fast camera were made with a THGEM foil without using a decoupling resistor in a Ne-CO$_2$-N$_2$ (90-10-5) mixture. On figure \ref{fig_12}, the optical recording of a primary discharge and the delayed DP is shown for a time interval lasting approximately 60~$\mu$s. The second frame in figure \ref{fig_12} shows the primary discharge at t=0~s. The delayed DP can be seen 48~$\mu$s after the primary discharge. The recorded frames in the time period between the primary and the delayed DP indicate that the primary discharge diminishes over some time period and just as it vanishes a small glow spot becomes visible in the vicinity of the THGEM bottom electrode. The glow remains visible until the delayed DP occurs. The delayed DP filament in all measurements originated from the glowing spot on the THGEM bottom electrode.

\begin{figure}[!htb]
\begin{center}
\includegraphics[width=1.0
\columnwidth]{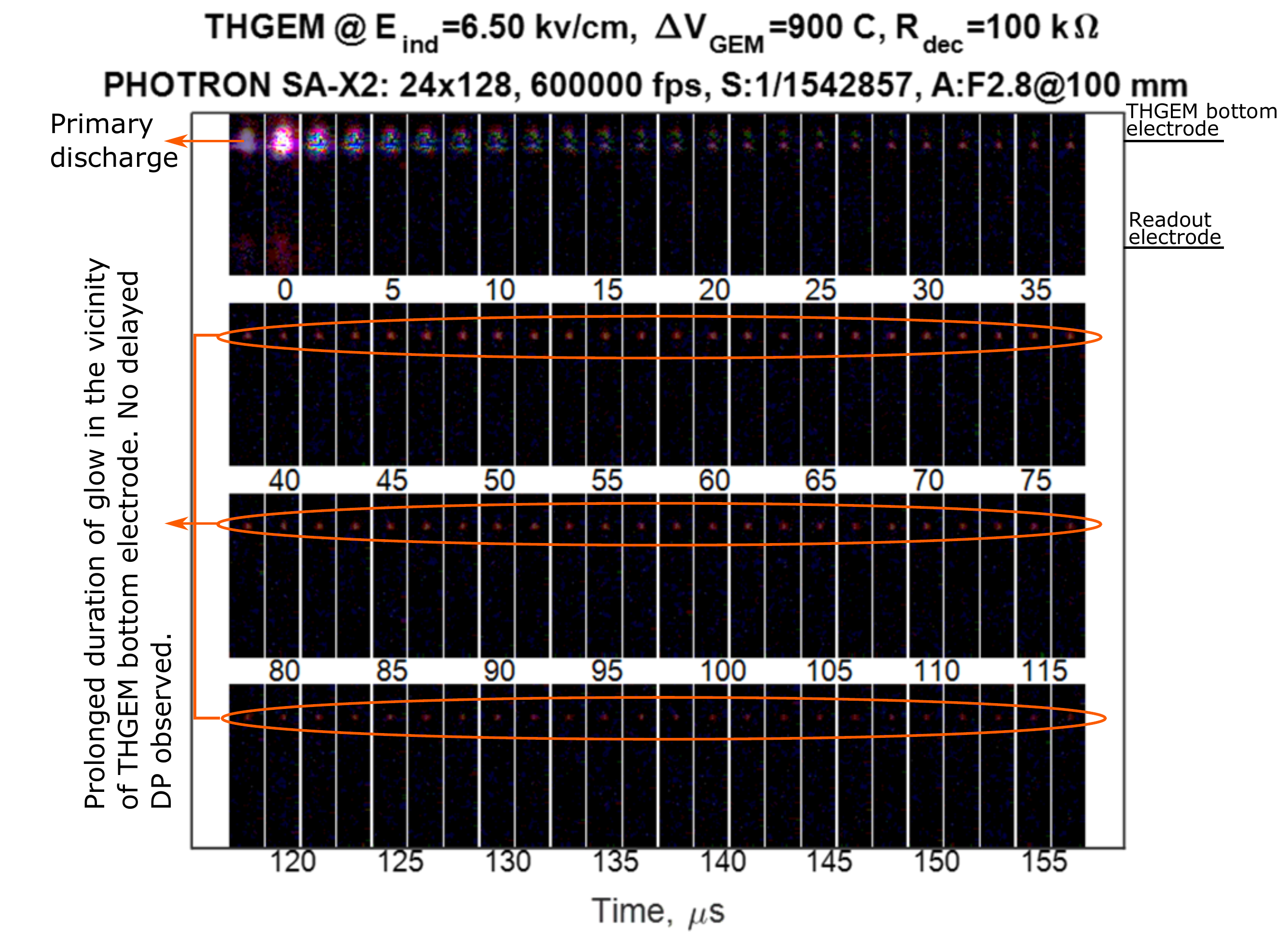}
\caption{A fast camera recording of the delayed DP with the THGEM foil at induction field value of 6.5 kV/cm using  100 k$\Omega$ decoupling resistor in a Ne-CO$_2$-N$_2$ (90-10-5) gas mixture.}
\label{fig_13}
\end{center}
\end{figure}

The measurements performed with a 100 k$\Omega$ decoupling resistor, figure \ref{fig_13}, show that for an induction field larger than the onset field for a delayed DP, a prolonged glow (lasting $\approx$~100~$\mu$s) in the vicinity of the THGEM bottom electrode can be observed. 

\begin{figure}[!htb!]
\begin{center}
\includegraphics[width=1.0
\columnwidth]{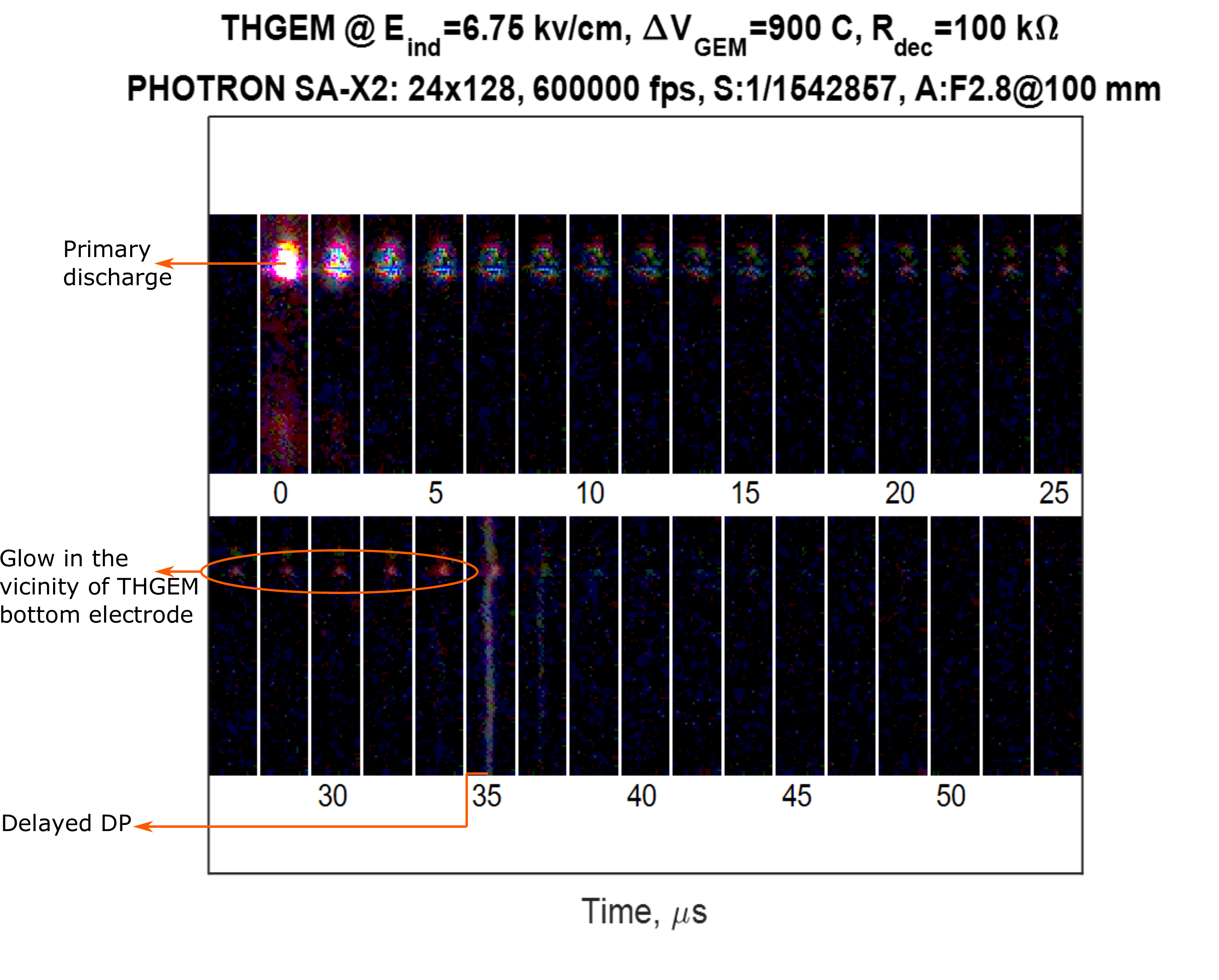}
\caption{A fast camera recording of the delayed DP with the THGEM foil at induction field value of 6.75 kV/cm using 100~k$\Omega$ decoupling resistor in a Ne-CO$_2$-N$_2$ (90-10-5) gas mixture.}
\label{fig_14}
\end{center}
\end{figure}

Figure \ref{fig_14} shows an optical recording of the delayed DP when using a 100~k$\Omega$ decoupling resistor. The delayed DP propagation can be observed 35~$\mu$s after the primary discharge. Even when the powering configuration uses a decoupling resistor, a glow in the vicinity of the THGEM bottom electrode is visible before the delayed DP. The optical appearance of the delayed DP is lower in intensity when using a 100~k$\Omega$ decoupling resistor than when there is no decoupling resistor. This is in agreement with earlier SLR optical and electrical measurements.  

\begin{figure}[!htb!]
\begin{center}
\includegraphics[width=1.0
\columnwidth]{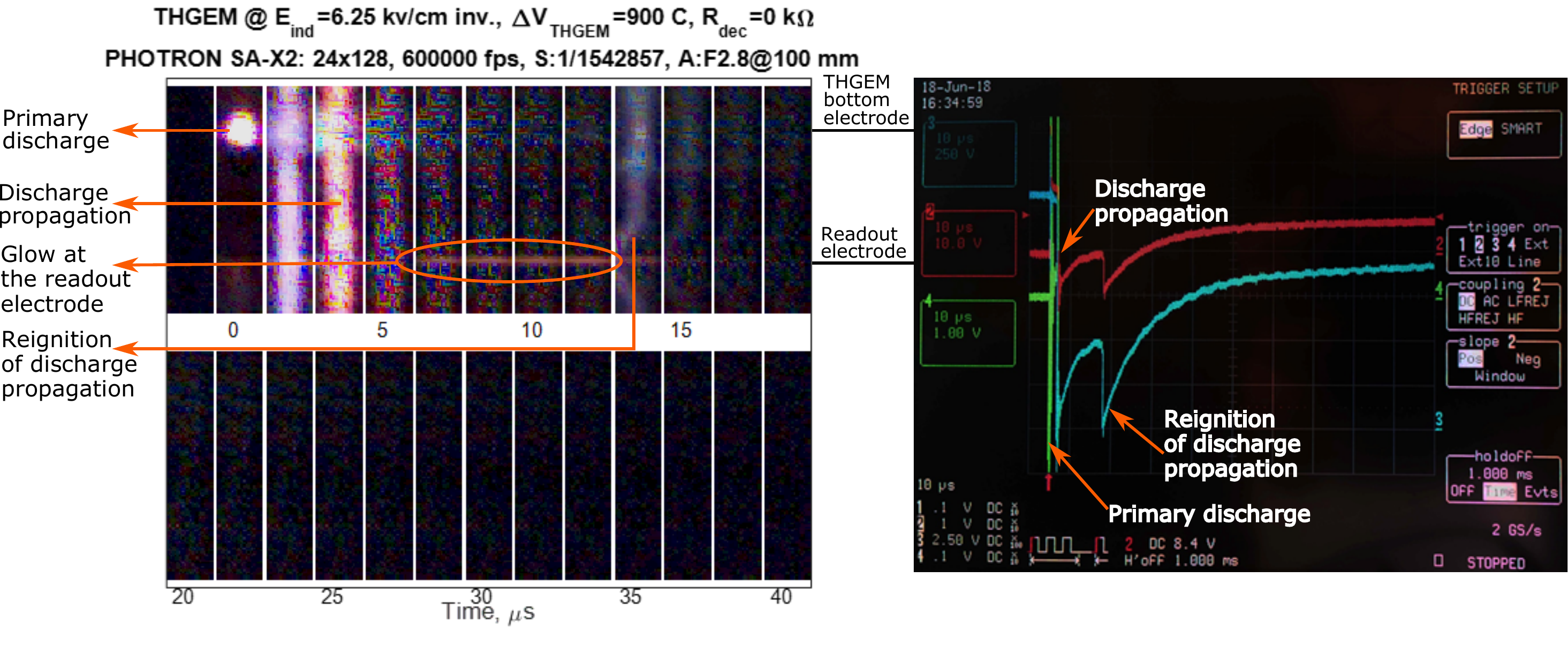}
\caption{Simultaneous fast camera (left panel) and oscilloscope waveform (right panel) recordings with the THGEM foil at an inverted induction field value of 6.25 kV/cm in a Ne-CO$_2$-N$_2$ (90-10-5) mixture. A reignition of the DP is visible 13.5 $\mu s$ after the primary discharge. Red signal on oscilloscope waveform (right panel) recordings represents voltage from THGEM top electrode while blue signal is voltage from THGEM bottom electrode and green signal is voltage on the readout capacitor.}
\label{fig_15}
\end{center}
\end{figure} 

\begin{figure}[h t b!]
\begin{center}
\includegraphics[width=1.0
\columnwidth]{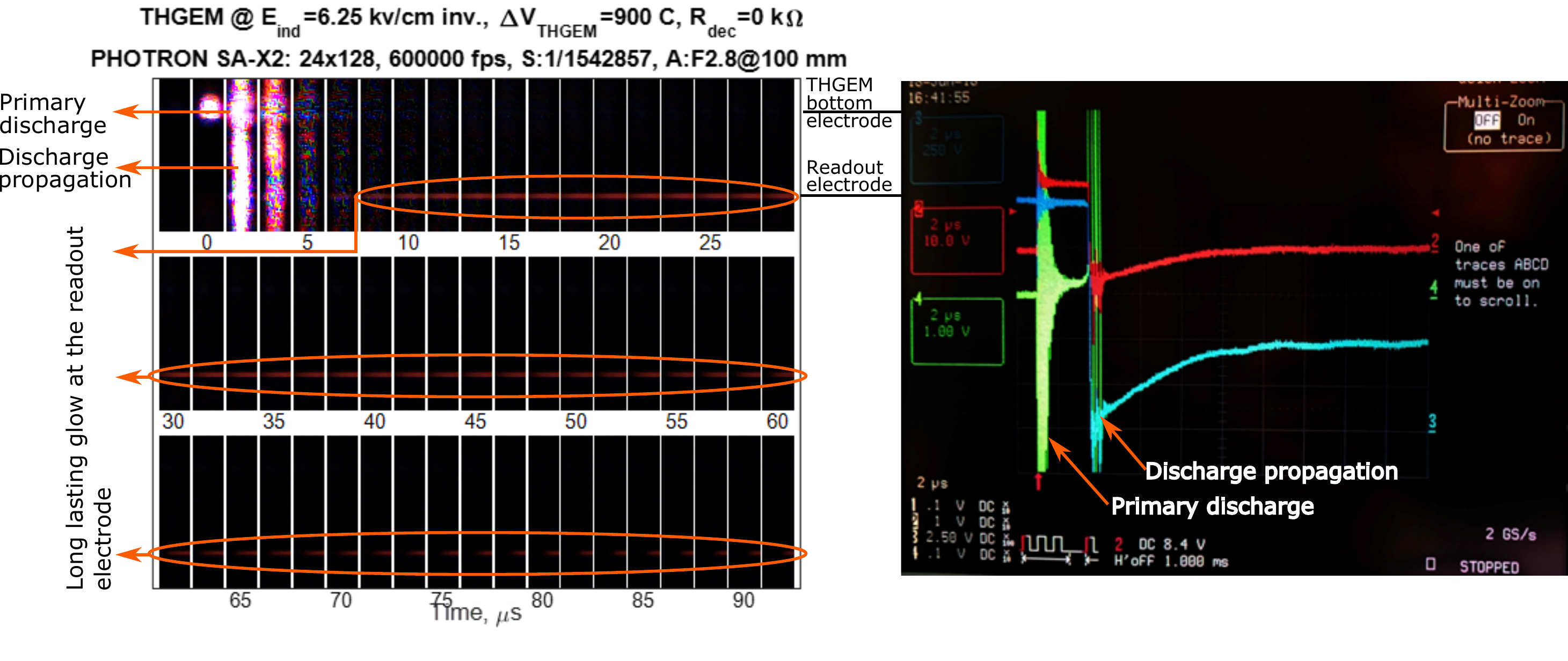}
\caption{Simultaneous fast camera (left) and oscilloscope waveform (right) recording with THGEM foil at inverted induction field 6.25 kV/cm  in a Ne-CO$_2$-N$_2$ (90-10-5) mixture. Long lasting glow is visible after the DP on the readout electrode. The oscilloscope channels colors correspond to the same signals as on figure \ref{fig_15}.}
\label{fig_16}
\end{center}
\end{figure} 

Next, we investigated the delayed DP with an inverted induction field. Figure \ref{fig_15} shows the fast camera measurements (left panel) and oscilloscope measurements (right panel) of the delayed DP at an inverted induction field value of 6.25 kV/cm. Approximately 3.5~$\mu$s after the primary discharge, the DP occurs. The optical appearance of the delayed DP with inverted induction field measurements is wider in diameter than with normal induction field measurements. Due to the geometry of the experimental set-up, the resolution had to be increased so it was not possible to record at frame rate larger than 600~000~fps. Therefore it was not possible to record more frames in the time period between the primary and the delayed DP. Figure \ref{fig_15} (left) shows that after the discharge propagation a glow is visible on the readout electrode which acts as a cathode due to the inverted field. It can also be observed that $\approx$ 13.5~$\mu$s after the primary discharge there is a re-ignition of the DP in the induction region. This is also visible on the recorded waveforms from the THGEM electrodes and the readout electrode, as shown on the right panel of figure  \ref{fig_15}.

Figure \ref{fig_16} shows a measurement in which on the second frame right after primary discharge, a delayed DP is observed. In this case, there was no reignition of the DP, only a long-lasting glow ($\approx$ 90 $\mu$s) on the readout electrode.

\subsection{High speed camera measurements with a  LP GEM foil}

\begin{figure}[h t b!]
\begin{center}
\includegraphics[width=1.0
\columnwidth]{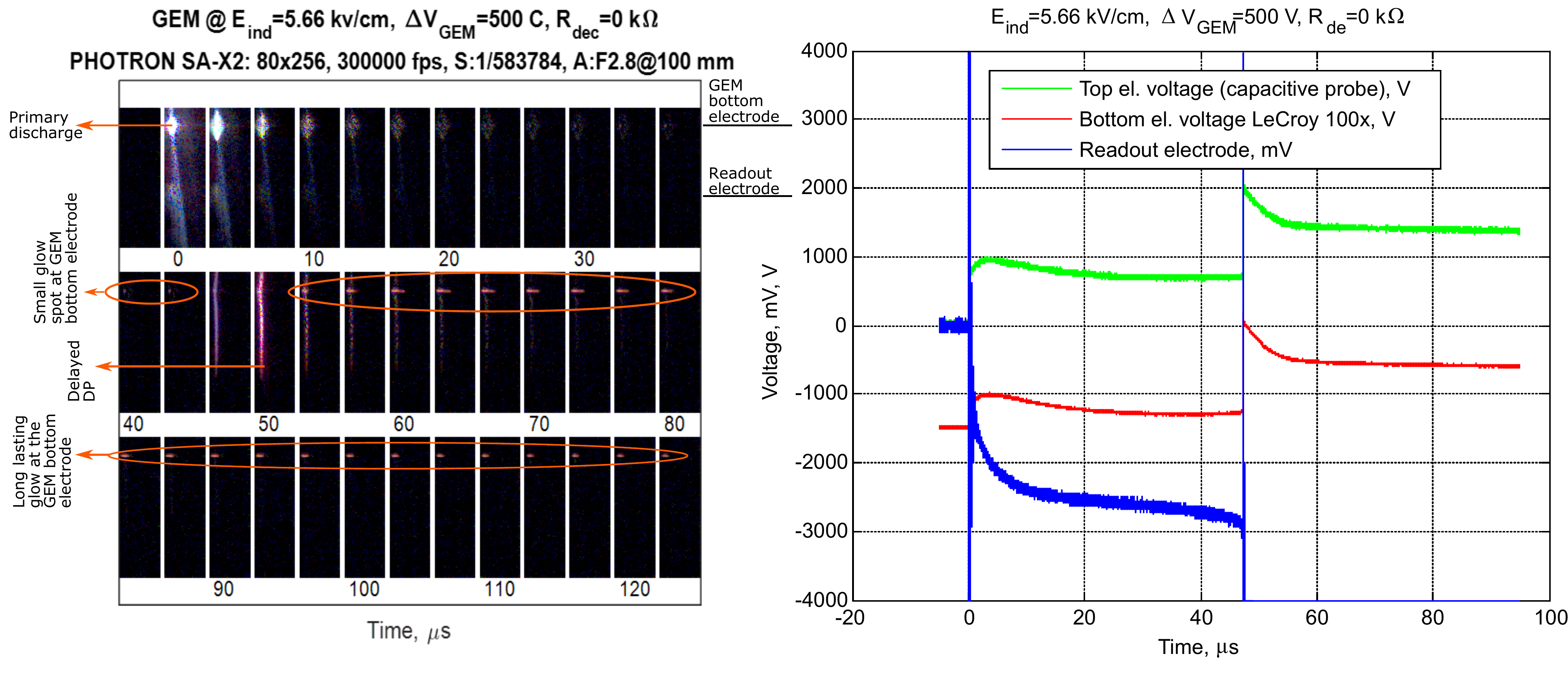}
\caption{Simultaneous fast camera (left)  and waveform recording (right) of the delayed DP with standard LP GEM foil at induction field 5.66 kV/cm  in a Ne-CO$_2$-N$_2$ (90-10-5) mixture.}
\label{fig_17}
\end{center}
\end{figure}

LP GEM foil measurements were repeated using a high-speed camera. Due to the fact that the GEM foil has a large number of holes (so the exact GEM hole where the primary discharge happens was unknown), it was necessary to increase the camera resolution, which resulted in a maximum recording frame rate of 300 000 fps. The left panel of figure \ref{fig_17} shows an optical recording of the delayed DP without a decoupling resistor at an induction field strength 5.66~kV/cm. The right panel shows the recorded voltages on the GEM electrodes and the readout electrode.  At the t=0~s frame, the primary discharge is visible as a plasma bulb under the GEM bottom electrode. The next frame shows that the primary discharge expands in volume. After that the intensity of the plasma bulb diminishes with time and $\approx$ 40~$\mu$s after the primary discharge, a small glow spot in the vicinity of the GEM bottom electrode becomes visible. A delayed DP occurs 47~$\mu$s after the primary and is connected to the small glow spot visible earlier.  Similarly to the measurements with the THGEM foil, after the delayed DP, a long-lasting glow can be observed near the GEM bottom electrode, suggesting the same physical mechanism responsible for delayed DP in GEM and THGEM, despite their geometrical differences. 

\begin{figure}[h t b!]
\begin{center}
\includegraphics[width=1.0
\columnwidth]{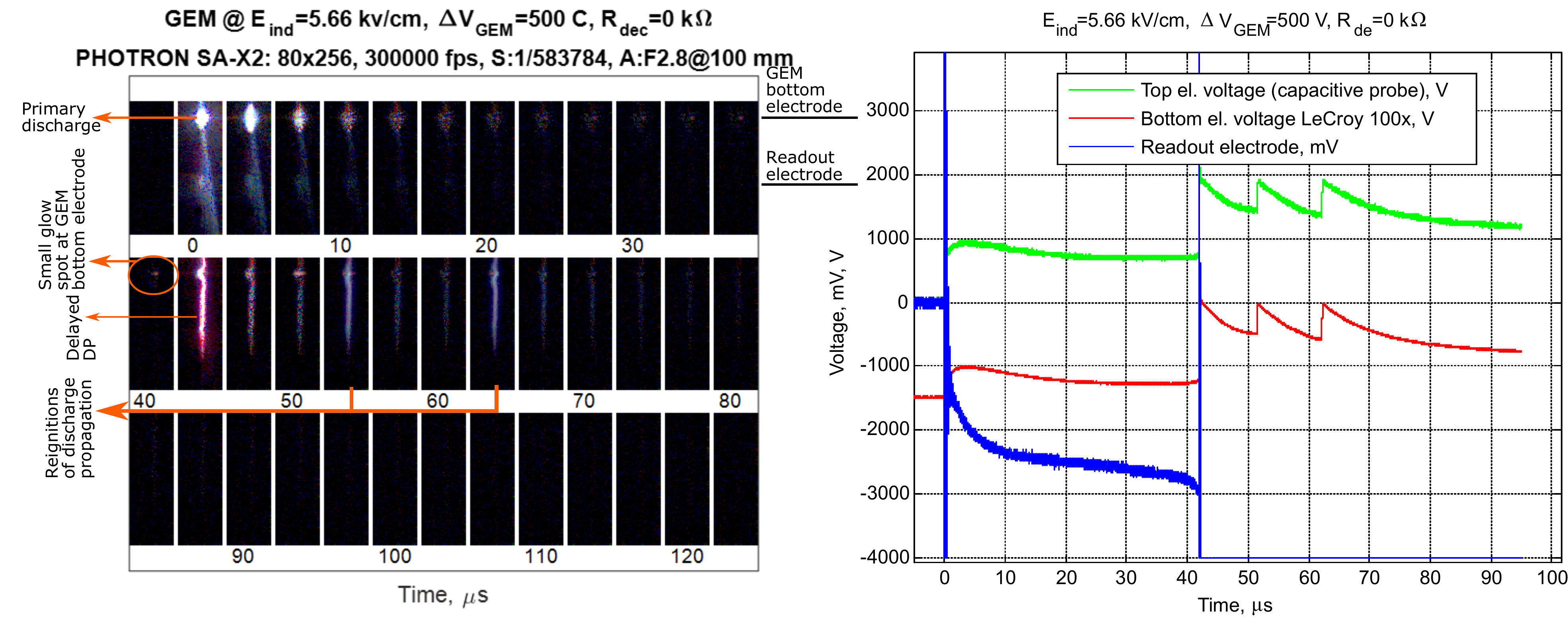}
\caption{Simultaneous fast camera (left)  and waveform recording (right) of the delayed DP with two reignitions using standard LP GEM foil at induction field 5.66 kV/cm  in a Ne-CO$_2$-N$_2$ (90-10-5) mixture.}
\label{fig_18}
\end{center}
\end{figure}

Both figure \ref{fig_17} and \ref{fig_18} show a delayed DP with the same induction field value of 5.66~kV/cm.  Two re-ignitions of the delayed DP can be seen in figure \ref{fig_18}.  A small glow is visible in the vicinity of the GEM bottom electrode before a delayed DP occurs and before the reignitions.  

\begin{figure}[h t b!]
\begin{center}
\includegraphics[width=1.0
\columnwidth]{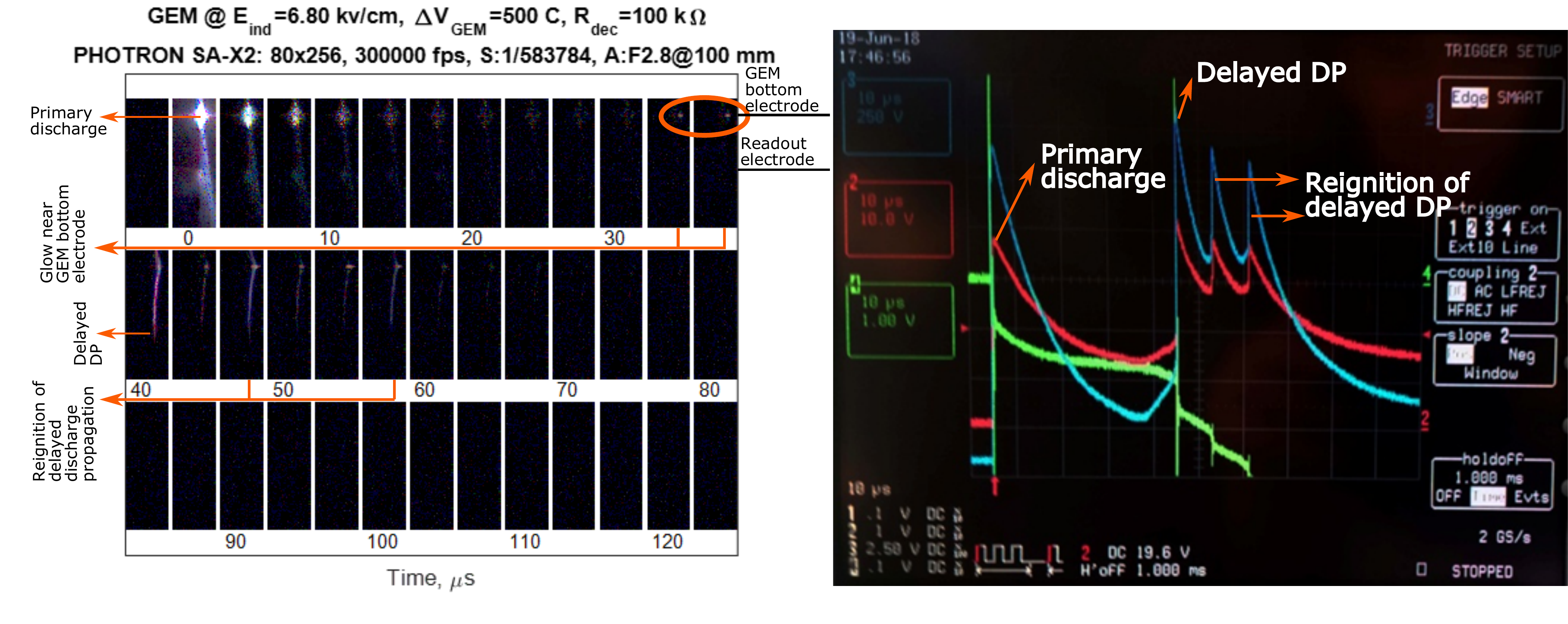}
\caption{Simultaneous fast camera (left)  and waveform recording (right) of delayed DP with standard LP GEM foil using 100 k$\Omega$ decoupling resistor at induction field 6.8 kV/cm  in Ne-CO$_2$-N$_2$ (90-10-5) mixture. The oscilloscope channels colors correspond to the same signals as on figure \ref{fig_15}.}
\label{fig_19}
\end{center}
\end{figure} 

Optical measurements with a 100~k$\Omega$ resistor are shown in the left panel figure \ref{fig_19}. Even when using a decoupling resistor, a glow in the vicinity of the GEM bottom electrode is visible right before the delayed DP, which occurs 40~$\mu$s after the primary discharge. The effect of re-ignition of a propagating discharge is observed after the delayed DP. Electrical measurement shown in the right panel figure \ref{fig_19} are in agreement with the recorded optical measurements. The voltage on the readout capacitor shown as a green signal in figure  \ref{fig_19} (right panel) indicates that there is a constant current flow through the induction region before the delayed DP and in between re-ignitions.

\section{Discussion}

Multiple physical mechanisms have been suggested as a possible explanation of the delayed DP. Photoelectrons from the drift electrode or electrons from the GEM top electrode created due to the slow collection of ions from either the alpha track or due to the ions created in the primary discharge have been considered as a possible charge source responsible for the delayed DP \cite{peskov2009research},\cite{sauli2016}. In order to obtain more information regarding the possible charge source responsible for the delayed DP, detailed electrical and optical measurements have been performed on a custom made measurement setup.

The current in the induction region after the primary discharge was recognized as a signal of interest and might indicate the cause of the delayed DP. 
The observed negative voltage build-up on the readout electrode capacitor confirms that there is a charge transfer in the time interval between the primary and the delayed DP. This voltage was used to determine the current in the induction region. It was necessary to obtain a smooth fit function of the recorded waveform even though there was high noise in the recorded voltage signal that originated from sparking. The current in the induction region was calculated by multiplying the time derivative of the voltage fit function with the capacitance value of 100~nF. The current through the 100~k$\Omega$ discharge resistor connected to the capacitor in parallel was not taken into account.

\begin{figure}[htb!]
\begin{center}
\includegraphics[width=0.99
\columnwidth]{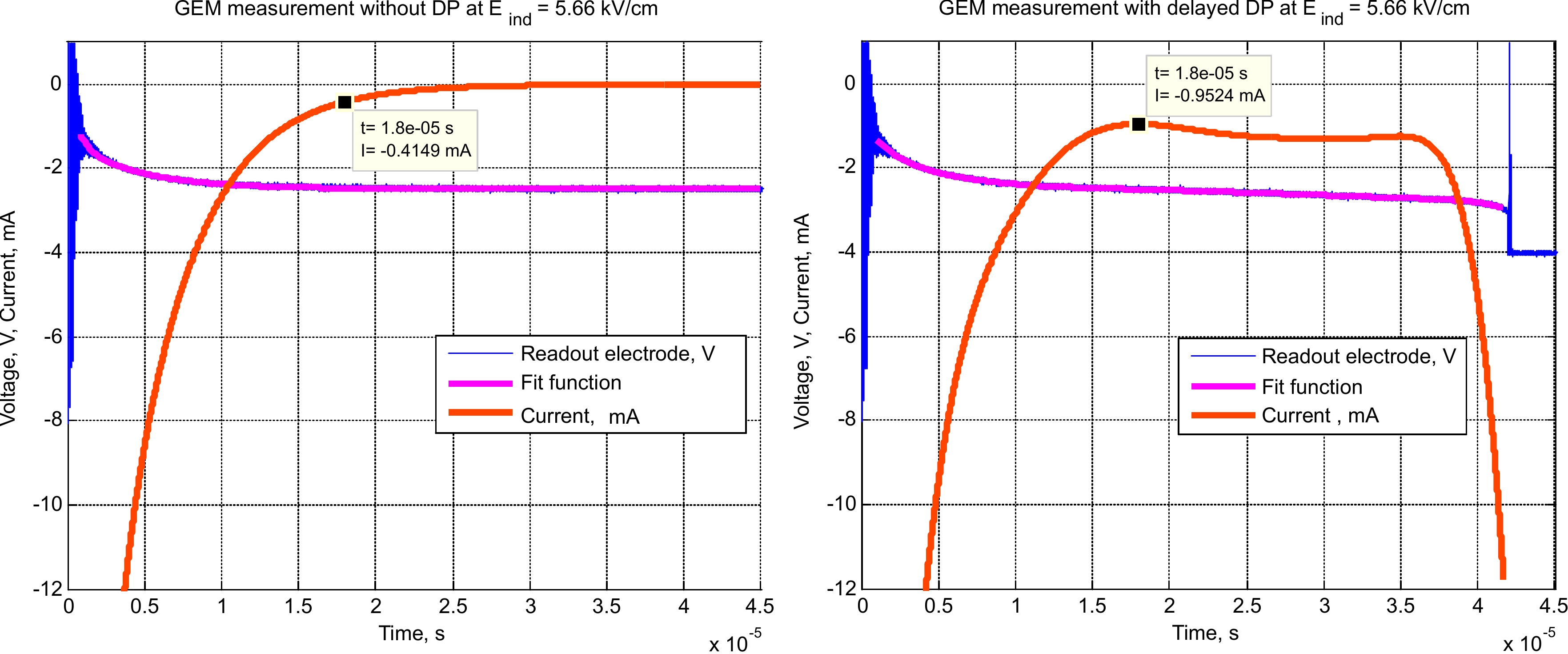}
\caption{The recorded signal on the 100 nF capacitor connected to the readout electrode (blue curve) with the obtained fit function (magenta curve) and the calculated current (orange curve) for GEM foil measurements with no delayed DP (left panel) and with the delayed DP (right panel) at an induction field value of 5.66 kV/cm.}
\label{fig_current_fit}
\end{center}
\end{figure}

Figure  \ref{fig_current_fit} shows an example of the calculated current (orange waveform) in the induction region for the measurements with the GEM foil at an induction field value of 5.66 kV/cm in an event with a delayed DP (right panel) and no delayed DP (left panel). The waveform shown on the right panel of figure \ref{fig_current_fit} corresponds to the event recorded with the high-speed camera shown on figure \ref{fig_18}.

The comparison of the delayed DP event and no DP event for the same induction field strength reveals a difference in the induction current behavior.  The calculated current for the event with no delayed DP event shown on the left panel of figure \ref{fig_current_fit} suggests that right after primary discharge there is an initial current spike of tens of milliamperes that decays to zero within 30~$\mu$s. 
On the other hand, the calculated current in the event with a delayed DP shown on the right panel of figure \ref{fig_current_fit} shows that after the initial current decay from the primary discharge, the
induction current does not decay to zero but reaches a minimum value of nearly 1~mA 18~$\mu$s after the primary discharge. After reaching its minimum value, the induction current remains almost constant during the next 18~$\mu$s. The current starts to increase rapidly 36.6~$\mu$s after the primary discharge and is followed by the delayed DP to the readout board after a couple of microseconds. It can be noticed that in the case of no delayed DP event there is no constant current regime indicating that the source of this constant current is responsible for the delayed DP.

\begin{figure}[htb!]
\begin{center}
\includegraphics[width=0.69
\columnwidth]{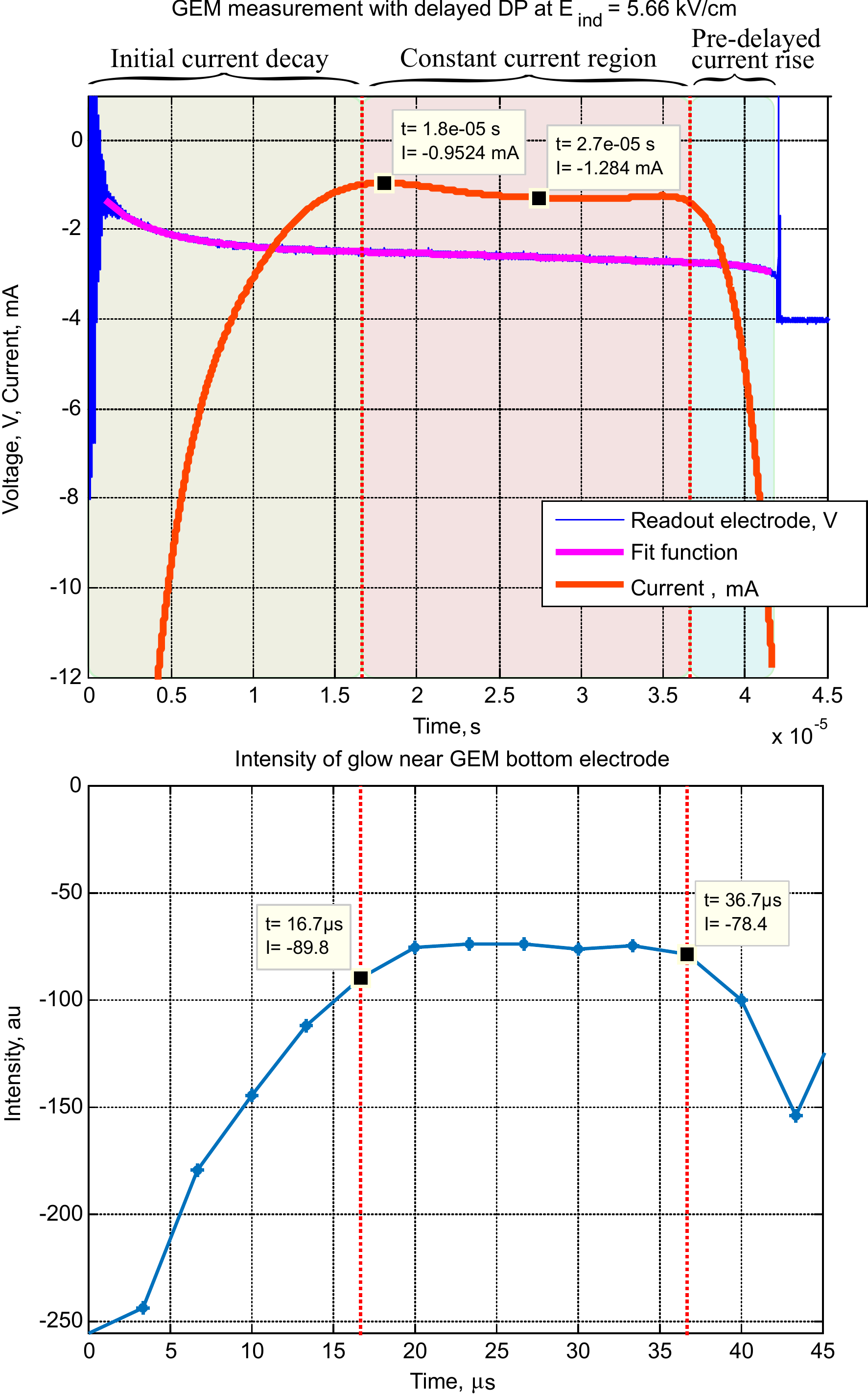}
\caption{Comparison of calculated current in induction region (up) and glow intensity near the GEM bottom electrode for measurements with GEM foil at induction field value of 5.66 kV/cm.}
\label{fig_current_intensity}
\end{center}
\end{figure}

The current behavior can be related to the optical recording from the high-speed camera in the left panel of figure \ref{fig_18} in order to obtain more information regarding the charge source in the induction region.  As noticed earlier, after a decrease in the intensity of the primary plasma bulb, a small orange glow becomes visible in the vicinity of the GEM bottom electrode, after which the delayed DP occurs and is connected to the glow.  The change in the glow intensity can be compared to the change of the calculated current in the induction region. The intensity of pixels located at the glow position was calculated for each recorded fame. When comparing the obtained light intensity shown in the bottom panel of figure \ref{fig_current_intensity} with the calculated current shown in the top panel of figure \ref{fig_current_intensity}, it can be seen that there is a similar trend in their behavior.
The bottom panel of figure \ref{fig_current_intensity} shows that at t= 0~$\mu$s the light intensity is maximum due to the primary discharge and then decreases for approximately 20~$\mu$s. After the intensity of the plasma bulb decreases it remains practically constant during the next 16~$\mu$s. Just one frame before delayed DP occurs, an increase in intensity can be observed once again. This leads to the conclusion that the constant current regime and the pre-delayed current increase is related to the observed GEM bottom glow, indicating that the GEM bottom electrode is responsible for the delayed DP.

The electrical measurements show three different current regimes in the induction region after the primary discharge: the initial current decay, a constant current regime, and a pre-delayed current rise, as shown in up panel of figure \ref{fig_current_intensity}. This behavior indicates that there are different contributions to the current in the induction region in the time between the primary and the delayed DP.

Optical measurements reveal valuable information regarding the time evolution of the delayed DP. The high-speed camera recordings show that after the primary discharge there is a plasma bulb formed below the (TH)GEM hole that does not cool off instantly and in the GEM measurements, it even expands in volume for approximately 3~$\mu$s after the primary discharge. The light emitted by the heated gas then slowly decreases in the intensity over the next 10-20~$\mu$s. After the intensity of the light from the primary plasma bulb decrease, an orange glow can be observed in the vicinity of GEM bottom electrode (cathode of the induction region). The delayed DP, connected to the glowing spot, is observed a couple $\mu$s after this decrease.

The initial current decay can be related to the bombardment of the GEM bottom electrode by the ions that are left over after the primary discharge. During the primary discharge, the gas in the vicinity of the GEM hole is heated to the large temperature and can heat the GEM bottom electrode up to the point where the electrode work function is significantly reduced. With an applied induction field this can lead to the emission of thermionic electrons from the bottom electrode, seen as a glow between primary and the delayed DP and a constant current regime in electrical measurements. The microscopic studies done on the single hole GEM foils show large deterioration of the electrode material as a consequence of the primary discharges \cite{merlin2018}. This is in accordance with our hypothesis of a heated bottom electrode. The thermal effects also explain the dynamics of the delayed DP with a rather high delay between the discharges. The optical measurements show an increase in the glow intensity before the delayed DP. This is also accompanied by a rise in the current. This can be explained with the thermal instability of the glow discharge. At higher induction fields, the constant current value and the glow intensity are higher,  heating up the gas and the electrode to the point where the thermal runaway triggers delayed DP much sooner.


\section{Conclusion}
The delayed DP to the readout electrode is one of the most threatening types of discharges in the GEM detector that can destroy sensitive readout electronics and compromise the detector stable operation. The main motivation of this work was to improve the understanding of the delayed DP mechanism.

Correlated optical and electrical measurements of delayed DP phenomena in GEM detectors were made with a single hole THGEM and a standard GEM foil. A robust single hole THGEM was used in order to preserve GEM foils and to be able to have a controlled primary discharge position.  It was observed that the delayed DP occurs even without a radiation source if an overvoltage is applied to the THGEM foil. This eliminates the possibility that the ions from alphas trigger the delayed DP. Electrical measurements performed on a GEM and the THGEM show that there is a current flow through the induction region between the primary and the delayed DP that can last for tens of microseconds. The current in the induction region appeared in three typical regimes. The regime of constant current was identified as an indicator of the delayed DP. 

Due to the relatively small time scales between the primary and the delayed DP, optical measurements were performed with a high-speed camera.  The induction region current value and the light intensity below the GEM hole were found to be correlated.  During the constant current regime, a glow at the GEM bottom electrode was formed. The light from the glow increases in intensity before the delayed DP event at the same time when the pre-delayed DP current rise is observed. The observed glow indicates that the charge source responsible for the delayed DP comes from the GEM bottom electrode.

A possible explanation for the delayed DP, with a delay of tens of microseconds, with respect to the primary, is the occurrence of the thermionic emission from the GEM bottom electrode due to the applied induction field and the heated electrode from the primary discharge. This either indicated as a glow between the primary and the delayed DP or as a constant current regime in the electrical measurements. Constant current can heat up the induction region cathode (GEM bottom electrode) up to the point where a thermal runaway triggers the delayed DP. These thermal effects could explain the microsecond scale of the time delay between the primary and the delayed DP.


\section*{Acknowledgements}
We want to thank the ALICE TPC Upgrade group and the RD51 collaboration members for useful discussions while conducting this research. 
We also want to thank Dewesoft d.o.o. for their contribution in acquiring high speed camera. This work has been fully supported by the Croatian Science Foundation under the project   IP-2018-01-4108.


\vspace{0.5 cm}
  \bibliographystyle{elsarticle-num} 
  \bibliography{literatura}

\end{document}